# ChASE: Chebyshev Accelerated Subspace iteration Eigensolver for sequences of Hermitian eigenvalue problems

JAN WINKELMANN, RWTH Aachen University, Germany

PAUL SPRINGER, RWTH Aachen University, Germany

EDOARDO DI NAPOLI*, Forschungszentrum Jülich, Germany and RWTH Aachen University, Germany

Solving dense Hermitian eigenproblems arranged in a sequence with direct solvers fails to take advantage of those spectral properties which are pertinent to the entire sequence, and not just to the single problem. When such features take the form of correlations between the eigenvectors of consecutive problems, as is the case in many real-world applications, the potential benefit of exploiting them can be substantial. We present ChASE, a modern algorithm and library based on subspace iteration with polynomial acceleration. Novel to ChASE is the computation of the spectral estimates that enter in the filter and an optimization of the polynomial degree which further reduces the necessary FLOPs. ChASE is written in C++ using the modern software engineering concepts which favor a simple integration in application codes and a straightforward portability over heterogeneous platforms. When solving sequences of Hermitian eigenproblems for a portion of their extremal spectrum, ChASE greatly benefits from the sequence's spectral properties and outperforms direct solvers in many scenarios. The library ships with two distinct parallelization schemes, supports execution over distributed GPUs, and it is easily extensible to other parallel computing architectures.

---

*Corresponding author

---

Article based on research supported by the Excellence Initiative of the German federal and state governments and the Jülich Aachen Research Alliance – High-Performance Computing. Financial support from the Deutsche Forschungsgemeinschaft (DFG) through grant GSC 111 is gratefully acknowledged.

Authors' addresses: Jan Winkelmann, RWTH Aachen University, AICES, Schinkelstraße 2, Aachen, 52062, Germany, winkelmann@aices.rwth-aachen.de; Paul Springer, RWTH Aachen University, AICES, Schinkelstraße 2, Aachen, 52062, Germany, springer@aices.rwth-aachen.de; Edoardo Di Napoli, Forschungszentrum Jülich, Jülich Supercomputing Centre, Wilhelm-Johnen-Straße, Jülich, 52425, Germany, RWTH Aachen University, AICES, Schinkelstraße 2, Aachen, 52062, Germany, e.di.napoli@fz-juelich.de, dinapoli@aices.rwth-aachen.de.

---

# 1 INTRODUCTION

The solution of a set of dense Hermitian eigenproblems organized in a sequence plays an important role in several scientific computing applications (e.g. condensed matter [26], thermoacustics [41], optoelectronics [47]). To solve for a portion of the exterior spectrum of such a sequence of problems, the expert computational scientist can choose among a large number of eigensolvers [2, 8, 24, 31]. In general, these eigensolvers are employed without any knowledge of the properties of the problems nor of their origin. As such, they solve each problem of the sequence in complete isolation and do not exploit possible features of the problems which are an intrinsic hallmark of being part of a sequence. By leveraging on subspace iteration [5, 38], a well-known and well-established iterative algorithm [4, 6, 28, 53], we present an alternative strategy and introduce the Chebyshev Accelerated Subspace iteration Eigensolver (ChASE) library. When tackling sequences of Hermitian eigenproblems, ChASE takes advantage of the distinctive features connecting adjacent problems in a sequence. At the core of the library, the Chebyshev filter is optimized to minimize the number of FLOPs necessary to declare eigenpairs converged. Thanks to the modern design of its interface, ChASE can be easily integrated in application codes and executed in parallel with a scalability comparable to, or better than, the state-of-the-art direct solvers even when used on isolated problems. We provide two parallelization approaches as part of the library and show how ChASE can be ported on a variety of parallel architecture with a moderately small effort.

When solving for an algebraic eigenvalue problem, the expert computational scientist can choose among a large number of eigensolvers encoded in standard packages. The selection of the best solver depends on several factors. Among them the size and sparsity of the matrices as well as the number and position of the required eigenpairs play a major role. For instance, large and dense eigenproblems are typically solved using a direct eigensolver unless only very few eigenpairs at the extreme ends of the spectrum are required, in which case an iterative eigensolver may be preferred. Therefore, the favored algorithm best compromises among many competing numerical properties so as to maximize performance while maintaining a good level of accuracy.

While the winning strategy to select the best algorithm is usually based on experience, existing eigensolver libraries are designed according to the principle of "separation of concerns": the library user is isolated from the intricacies of the implementation and the library developer is oblivious of the scientific application the eigenproblem is emerging from. The interaction between user and library is guaranteed by a well designed interface which standardize its usage. On the up side, this so called "black-box" approach allows for the universal use of a library. On the down side, no extra information, related to the specific scientific domain from which the eigenproblem originates, is exploited by the selected eigensolver.

We propose to step away from the black-box approach and address a class of eigenproblems exhibiting a certain type of well-defined properties. Such a strategy implies that the resulting library is tailored to a specific class of applications, and loose generality of usage. On the other hand, the solver takes advantage of the problem properties and becomes very efficient. With this general strategy in mind, we designed ChASE, a sophisticated library based on the subspace iteration algorithm with Chebyshev acceleration, which targets extremal eigenpairs of dense Hermitian eigenproblems. Specifically, ChASE performs at its best when solving sequences of eigenvalue problems where adjacent problems possess a certain degree of correlation. In addition, ChASE takes advantage of those cases where, instead of full accuracy, is sufficient to solve each eigenproblem with a level of accuracy that depends on the sequence index. A typical example of such applications is Density Functional Theory (DFT) [26] where the solution to a non-linear partial differential equation is tackled by generating and solving tens to hundreds of linear eigenvalue problems in a self-consistent fashion over dozens of iterations. Similarly, any non-linear eigenvalue problem solved by the method of successive linearization [37] gives rise to sequences of correlated algebraic eigenproblems that are the target of ChASE.

ChASE exploits the properties of eigenproblem sequences in two ways. Correlation between eigenproblems in a sequence is often expressed as an increasing collinearity between their respective eigenvectors. Because it is based on the subspace iteration algorithm, ChASE can receive as input as many vectors as the desired ones. By using the solution of a problem in a sequence as starting vectors for the next problem, ChASE can experience up to 3.5× speedups [6]. To work efficiently, polynomial acceleration of subspace iteration needs to have accurate estimates bounding the sought after spectrum from above and below. Using the eigenspectrum solving for a problem as an approximation for the next one eliminates the need for computing the spectral bounds, further accelerating the solution of each problem and the sequence overall.

At the algorithmic level, our original contributions to ChASE revolve around the re-design of two of the most important computational tasks: spectral bounds estimation, and the Chebyshev filter. We increased the accuracy of the spectral bounds when ChASE is used in isolation. This goal is achieved by estimating the spectral density [29] through the manipulation of few repeated Lanczos steps. Accurate evaluations for the bounds are obtained by carefully tuning the value of the parameters controlling the density approximation. We optimized the polynomial degree of the Chebyshev filter and reduced up to 20% the number of FLOPs necessary to declare each of the eigenpairs converged. Such feat is achieved by a careful computation of the convergence ratio for each filtered vector which is then used to determine the corresponding minimal polynomial degree. In addition, the library is templated for single and double precision as well as for real symmetric and complex Hermitian matrices. Moreover, we provide the ability to offload the bulk of the computation to GPU accelerators.

Among the state-of-the-art black-box eigensolver libraries for distributed memory parallelism, we can distinguish two main paradigms. On the one hand, there are libraries that belong to monolithic packages, e.g. Anasazi [2] as part of Trillinos [21], or SLEPc [20] as an extension to PETSc [3]. These libraries are based on the software structure provided by their respective packages and are at their best when the application code is written entirely in Trillinos or PETSc primitives. Because they deliver a large set of features, the underlying packages offer a powerful framework, but may lack flexibility of integration with software running on specialized computing architectures (e.g. GPU clusters), and using optimized low level kernels. On the opposite side of the fence, there are stand-alone libraries without large dependencies like the ARPACK [27] or the more modern PRIMME package [43]. These packages are more selective in the functionalities they provide and are usually more flexible and easier to integrate in any application code independently from data distribution, the choice of low-level kernels, and the computing platform. The ChASE library realizes a modern version of the latter strategy.

We provide a stand-alone high-performance parallel implementation of ChASE that not only implements our algorithmic contributions but also promises (1) portability to heterogeneous architectures and (2) easy integration into existing codes. We achieve our goal by separating the implementation of the ChASE algorithm from the required numerical kernels via an interface based on a pure C++ abstract class. Classes derived from this interface handle data distribution and (parallel) execution of each kernel. The required numerical kernels are based on BLAS-3 [16] compatible kernels, such as a (parallel) matrix-matrix multiplication. This modern "stand-alone" strategy grants ChASE an unprecendented degree of flexibility which makes the integration of this library in most application codes quite simple. ChASE efficiently uses available machine resources.

The vast majority of FLOPs are spent on dense Hermitian matrix-matrix multiplications (HEMMs), which attain a significant percentage of peak performance on modern architectures. Matrix-matrix multiplications parallelize well, especially in weak-scaling regimes making ChASE perform well against other state-of-the-art solvers. On representative problems we out-perform Elemental's [35] state-of-the-art direct solver, and Anasazi [2].

The reminder of this paper is organized as follows. In Sec. 2 we give a rigorous definition of sequences of correlated eigenproblems and present some examples of their origin. The design of the ChASE algorithm together with its main tasks and parameters are introduced in Sec. 3. Sec. 4 illustrates our main original contributions to the algorithm: the degree optimization of the Chebyshev filter and the enhanced spectral estimates. In Sec. 5 we explain the structure of ChASE's parallel implementations and how to use the library. Numerical experiments and performance evaluations are presented in Sec. 6. Finally, Sec. 7 summarizes our work and gives an outlook on possible further developments.

## 2 SEQUENCES OF EIGENVALUE PROBLEMS IN SCIENTIFIC COMPUTING

A sequence of real numbers is usually defined to be an unlimited set of numbers ordered from smaller to larger in such a way that there exists a connection between two or more adjacent numbers of the sequence. Analogously to sequences of numbers, one can generalize the concept of *sequence* to an ordered set of square matrices. In general, the entries of a matrix in a sequence will not be related to the entries of previous matrices by a recurrence relation, so our definition of sequence has to be smarter. Since a square matrix implicitly defines an eigenproblem, we propose to use the matrix spectral properties as a more natural set of features to be used in a definition of a *sequence of matrices*. The idea of sequences of matrices can be naturally extended to include sequences of algebraic eigenproblems.

### 2.1 Sequences and correlation

A standard algebraic eigenproblem $P$ is defined as $A\hat{x} = \lambda\hat{x}$, where $A \in \mathbb{C}^{n^2}$ is a non-defective square matrix, while $\hat{x} \in \mathbb{C}^n \setminus \{0\}$ and $\lambda \in \mathbb{C}$ are respectively a vector and a number. In general there are $n$ of such $(\hat{x}, \lambda)$ pairs which are collected in a matrix $\hat{X} \doteq [\hat{x}_1, \ldots, \hat{x}_n]$ of eigenvectors and a vector $\Lambda \doteq [\lambda_1, \ldots, \lambda_n]$ of eigenvalues. In this work we exclusively address Hermitian eigenproblems with $A = A^H$, $\hat{X}^{-1} = \hat{X}^H$, and $\Lambda \in \mathbb{R}^n$, but the following definition can be equally used for more general problems.

*Definition 2.1.* A sequence of eigenproblems is defined as an $N$-tuple

$$\{P\}_N \doteq P^{(1)} \cdots P^{(\ell)} \cdots P^{(N)}$$

of distinct problems $P^{(\ell)}$ of identical size $n$ such that one or more spectral properties of the $\ell + 1$-problem are non-linearly related to one or more of the spectral properties of at least one of the previous problems in the sequence.

The definition above removes the need for a recurrence relation between the entries of the matrices in the sequence, and makes way for a general relation between objects such as the trace, the determinant, the spectral radius, etc. While the definition above implies a connection between matrices defining successive problems, it does not specify the nature of the relation nor which objects are related. For practical purposes we add to such a general definition the concept of *q-vector correlation*, which is based on the definition of canonical correlation [18] applied to the rank $q$ matrix of eigenvectors $\hat{Y}^{(\ell)} = [\hat{x}_1^{(\ell)}, \ldots, \hat{x}_q^{(\ell)}] \subset \hat{X}^{(\ell)}$ corresponding to the extremal eigenvalues of each problem.

*Definition 2.2.* The eigenproblems of a sequence $\{P\}_N$ are said to be $q$-vector correlated if $\forall \ell$ and $q < n$ it exists a non-negative constant $\delta^{(\ell)}$, with $\delta^{(\ell)} \leq \delta^{(\ell-1)}$ and $\delta^{(1)} < 1$, such that

$$\sigma_k^{(\ell)} = \max_{\substack{a \neq i_1, \ldots, i_{k-1} \\ b \neq j_1, \ldots, j_{k-1}}} \left[ \cos \angle \left( \hat{x}_a^{(\ell)}, \hat{x}_b^{(\ell-1)} \right) \right] \doteq \cos \angle \left( \hat{x}_{i_k}^{(\ell)}, \hat{x}_{j_k}^{(\ell-1)} \right),$$

and

$$1 - \sigma_k^{(\ell)} < \delta^{(\ell)} \qquad k = 1, \ldots, q.$$

The definition implies that in a *sequence of correlated eigenproblems*[1] the solutions of two adjacent problems are connected in such a way that the the angle between each of the corresponding eigenvectors gets progressively reduced. The vector correlation not only suggests a relation among successive eigenpairs, but it also indicates the tendency for such eigenpairs to be more closely coupled as the sequence progresses. Moreover, this definition lends itself to be characterized both in exact mathematics or in approximate numerics [15, 18]. Despite the restrictive nature of this definition, sequences of correlated eigenproblems emerge from many scientific applications.

## 2.2 Applications

Correlated eigenproblems appear in many fields of applied science. For instance, when the matrices defining the problems are the result of the discretization of a physical system or the decomposition of a specific domain [30], two correlated eigenproblems are the result of a small perturbation of the physical system or change of the domain [50]. From this perspective, sequences of eigenproblems can be as short as made by just two problems. A more interesting case is provided by the solution of non-linear over-damped Hermitian eigenvalue problems $T(\lambda)\hat{x} = 0$ where $T(z)$ is in general a non-polynomial function of $z$. Such problems are guaranteed to have $n$ real eigenvalues [36], and when solved by successive linearization [37] or safeguarded iteration [48], can lead to long sequences of correlated linear eigenproblems.

There are cases when the non-linear function $T$ also depends on the eigenvectors or a selection of them. In such cases, the non-linear eigenvalue problem can be written as $\left[ T(\Lambda, \hat{Y}) - \lambda \mathbb{I} \right] \hat{x} = 0$, where $\Lambda \in \mathbb{C}^{q \times q}$ and $\hat{X} \supset \hat{Y} \in \mathbb{C}^{n \times q}$. A popular ansatz for this class of problems is the use of self-consistent iterations: $T(\Lambda_0, \hat{Y}_0)$ is initialized with a reasonable guess for the eigenpairs $(\Lambda_0, \hat{Y}_0)$ and the resulting linear eigenproblem is solved with standard techniques. The new set of eigenpairs $(\Lambda_1, \hat{Y}_1)$ are then used to re-initialize $T$ in the hope of reaching self-consistency within an accepted margin of error. One of the typical examples of such non-linear eigenvalue problems is given by Density Functional Theory (DFT), which constitutes the "standard model" in atomistic condensed matter computations [26].

*2.2.1 A typical example: Density Functional Theory.* The success of DFT is based on the fundamental theorem of Hohenberg and Kohn [22] which states that given the electronic Hamiltonian $H$ describing a multi-atoms quantum mechanical system, there exists a functional $E[\mathfrak{n}] = \frac{\langle \Psi | H | \Psi \rangle}{\langle \Psi | \Psi \rangle}$ such that $E_0 = \min_{\mathfrak{n}} E[\mathfrak{n}]$, with the eigenstates $|\Psi\rangle$ of $H$ being some function of the charge density $\mathfrak{n}$. Despite its simplicity, the theorem only sets the stage but does not provide a way to compute the functional. In its stead an equivalent problem is solved

$$\left( -\frac{\hbar^2}{2m} \nabla^2 + V[\mathfrak{n}] - \varepsilon_i \right) \phi_i = 0 \qquad \mathfrak{n} = \sum_{i=1}^{q} |\phi_i|^2$$

---

[1]From now on we drop the prefix *q-vector* in front of *correlation* and assume it implicitly.

In the language of linear algebra this can be written as $\left[T(\hat{Y}) - \lambda \mathbb{I}\right] \hat{x} = 0$. Typically physicists solve this problem self-consistently by starting with an educated guess for the charge density $\mathfrak{n}$, and iteratively compute a new density $\mathfrak{n}'$, until the distance between $\mathfrak{n}$ and $\mathfrak{n}'$ is below a certain threshold. At each iteration $\ell$ a linear algebraic eigenvalue problem $T(\hat{Y}^{(\ell-1)})\hat{x}^{(\ell)} = \lambda^{(\ell)}\hat{x}^{(\ell)}$ is solved which is typically correlated, in the sense of Def. 2.2, with the problem at iteration $\ell - 1$ [15]. The whole set of problems from beginning to end constitutes a classic example of a sequence of correlated eigenproblems. Examples of sequences from DFT will be used in Sec. 6.1 to illustrate how ChASE exploits the correlation.

*2.2.2 Exploiting the correlation.* Our definition of correlation is based on the definition of canonical correlations which are also known as the cosine of principal angles between subspaces [23] of the space spanned by $\hat{Y}^{(\ell)}$. When dealing with a sequence of correlated eigenproblems the angles get smaller as one travels along the sequence, implying that the corresponding subspaces become increasingly aligned. This simple observation suggests that the solution $\hat{Y}^{(\ell)}$ of problem $P^{(\ell)}$ can be accelerated by inputting the solution $\hat{Y}^{(\ell-1)}$ of the previous problem $P^{(\ell-1)}$ into the eigensolver. The selection of an appropriate method is paramount in order to maximally exploit approximate solutions. In an earlier work [14], we showed that subspace iteration with Chebyshev acceleration is the best candidate for such an approach. The Chebyshev acceleration works best if a good estimate of the interval to be filtered by the Chebyshev polynomials is provided. For sequences of correlated eigenproblems this information is automatically included in the spectrum of the $\ell - 1$ problem. The only missing element is an upper bound of the largest eigenvalue. We will show that such a bound can be easily extracted by a relatively inexpensive Lanczos procedure [52].

# 3 THE ORIGIN AND STRUCTURE OF THE CHASE LIBRARY

In order to understand the context that lead to the present work, we provide a brief historical review of Subspace Iteration (SI) from its earliest version up to current implementations.

## 3.1 The Path to a Modern Subspace Iteration Algorithm

SI is probably one of the earliest iterative algorithms to be used as a numerical eigensolver. The work by L. Bauer in 1957 [5] is arguably one of the the earliest articles in the scientific literature mentioning the application of SI to the solution of the symmetric algebraic eigenvalue problem. Several were the attempts to further develop and generalize SI in the 1960s and 1970s, when the Lanczos algorithm was still in its infancy. The first notable effort in this direction is the fundamental work of Rutishauser in a number of papers spanning from 1969 to 1970 [38, 39]. Rutishauser builds on Bauer's Simultaneous Iteration method and introduces many of the key ideas included in modern implementations of eigensolvers based on subspace iteration. He also presents a fairly robust complete implementation of its algorithm in a software library named *ritzit*.

After Rutishauser's first article, several authors contributed to improve the SI algorithm. For example, Stewart promoted the Ritz projection, which reduces the eigenproblem onto the active search space using a Jacobi step, to a full Rayleigh-Ritz step. Such an upgrade eliminates the problem of having to deal with non-positive definite matrices for non-definite Hermitian eigenvalue problems [44]. In addition, Stewart demonstrates that adding the Rayleigh-Ritz step to the orthogonal iteration enhances the convergence of the eigenpairs. Almost at the same time, civil engineers proposed a version of subspace iteration based on an algorithm very similar to Rutishauser's, but illustrated in a language and formalism somewhat different from conventional numerical analysts [10]. A clear review of the improved *ritzit* algorithm can be found in Parlett's book [32].

In parallel to these developments, SI was generalized by several authors to non-Hermitian and non-symmetric eigenproblems (see for example [45]). In the following two decades, the development of iterative eigensolvers for the Hermitian eigenvalue problem took on a different direction due to the revival of the Lanczos algorithm and its variations. Subspace iteration eigensolvers saw a resurgence in popularity starting in the middle of the 2000s with the application to electronic structure theory, first in the context of sparse eigenproblems [51, 53], and later also for sequences of dense eigenproblems [4, 6, 28]. Two are the main reasons for the come back: first, there was the emerging need to solve for the entire subspace without the need to resolve the single eigenpairs [53]. Second, and more pertinent to the current work, SI has the ability to receive approximate eigenvectors as input, and in doing so, considerably decrease its time-to-convergence [14]. The present work presents our research efforts in the latter direction.

Let us briefly mention the convergence properties of SI. For a detailed study including optimization of polynomial filter degree we refer the reader to the companion manuscript [13]. Subspace iteration is by definition a block solver since it attempts to build an invariant eigenspace by repeatedly multiplying a block of vectors $\hat{V}$ with the operator $A$ to be diagonalized. It is a known fact that, under the mild assumption that the matrix $X^H \hat{V}$ is invertible, any implementation based on subspace iteration—including a QR factorization step which eliminates the chance of rank deficient cases—converges linearly. The addition of a Rayleigh-Ritz procedure enhances the convergence of eigenvalues and eigenvectors in the case of Hermitian eigenproblems [44]. Let us define $\theta_a$ as the angle between $\hat{x}_a$ and span$(\hat{V} \leftarrow A^k \hat{V})$, then backward perturbation analysis shows that the Ritz vector $\hat{v}_a$ has a better bound for the linear convergence

$$\sin \angle (\hat{x}_a, \hat{v}_a) = O(\theta_a) \tag{1}$$

while the Ritz value $\tilde{\lambda}_a$ converges quadratically

$$\left| \tilde{\lambda}_a - \lambda_a \right| = O\left([\theta_a]^2\right)$$

([46] pp. 286-90).

Finally we would like to spend a few words on the influence of the algorithm originally proposed by Rutishauser on the present work. It is surprising to see that all major aspects of SI can already be found in his paper from 1969 [38], including insights on when SI should be used and how to determine a good set of initial input parameters. All the key ingredients like polynomial filtering, QR factorization, search space projection, and a Ritz iteration are already present. In addition, Rutishauser suggests:

(1) to use an initial Ritz iteration in order to have an estimate of the bounds of the eigenvalue interval to be filtered out;
(2) the idea of an initial iteration with a low polynomial degree to avoid large error in computing the extremal eigenvalues on the interval;
(3) to combine a limited number of filtering iterations before performing the orthonormalization as a compromise between minimizing the computation and avoiding the filtered vectors being excessively aligned to the dominant one;
(4) to carefully select the initial set of vectors since an unlucky choice can substantially influence convergence;
(5) to limit the maximum value of the polynomial degree used in the filter to mitigate the effect of floating point errors;
(6) a locking mechanism that stores converged eigenpairs and excludes them from further filtering.

Two main differences with respect to modern implementations exist: first, the Ritz iteration is performed on the product $\left(\hat{V}\right)^H \hat{V} = (R)^H R$, with $R$ coming from the QR factorization of $\hat{V}$. This expedient (known as Jacobi step) reduces the overall amount of floating point operations but is only applicable to positive definite $A$. Second, eigenpair residuals are used in the stopping criterion only after the angles $\theta_a$ undergo stagnation and are accepted as final. While this criterion ensures that the algorithm will terminate even when the requested accuracy cannot be achieved, it limits its flexibility when lesser accuracies are required.

### 3.2 The ChASE Algorithm

In the present work we illustrate a modernized version of Rutishauser algorithm including substantial algorithmic optimizations and stabilizations. The resulting algorithm, the Chebyshev Accelerated Subspace iteration Eigensolver (ChASE), is presented in Alg. 1. The colored lines of pseudocode distinguish our original contributions from the standard SI algorithm often presented in the literature. In this and following subsections we introduce the reader to the characteristics and subtleties of the standard SI algorithm. In Sec. 4 we explain the original contributions which promote the standard SI algorithm to the full ChASE algorithm.

---

**Algorithm 1** The ChASE algorithm: SI plus our original contributions

---

**Input:** Hermitian matrix $A$, number of desired eigenpairs nev, threshold tolerance for residuals tol, initial polynomial degree deg, search space increment nex, approx and optim flags, vector matrix $\hat{V} \equiv [\hat{v}_1 \ \dots \ \hat{v}_{nev+nex}]$ and estimates $\mu_1$ and $\mu_{nev+nex}$.

**Output:** nev extremal eigenpairs $(\Lambda, \hat{Y})$, with $\Lambda = [\lambda_1 \ \dots \ \lambda_{nev}]$ and $\hat{Y} \equiv [\hat{y}_1 \ \dots \ \hat{y}_{nev}]$, and their residuals $[\text{Res}(\hat{y}_1, \lambda_1) \ \dots \ \text{Res}(\hat{y}_{nev}, \lambda_{nev})]$

---

1: $m_{1:nev+nex} \leftarrow$ deg        ▷ INITIAL CONSTANT DEGREE
2: $(b_{sup}, \mu_1, \mu_{nev+nex}, \hat{V}) \leftarrow$ LANCZOS($A$, approx)        ▷ BOUND AND OPTIONAL INPUT

3: **while** size($\hat{Y}$) < nev **do**
4:      $\hat{V} \leftarrow$ FILTER($A$, $b_{sup}$, $\mu_1$, $\mu_{nev+nex}$, $\hat{V}$, $m$, optim)        ▷ USES ARRAY OF DEGREES
5:      $\hat{Q} \leftarrow$ ORTHOGONALIZE($[\hat{Y} \ \hat{V}]$)        ▷ QR ALGORITHM
6:      $\tilde{Q} \leftarrow [\hat{Q}_{:,\text{size}(\hat{Y})+1} \ \dots \ \hat{Q}_{:,nev+nex}]$        ▷ REDUCE TO ACTIVE SUBSPACE
7:      $(\hat{V}, \tilde{\Lambda}) \leftarrow$ RAYLEIGH-RITZ($A$, $\hat{Q}$)
8:      Compute the residuals Res($\hat{V}, \tilde{\Lambda}$)
9:      $(\hat{V}, \Lambda, \hat{Y}) \leftarrow$ DEFLATION $\mathcal{E}$ LOCKING($\hat{V}, \tilde{\Lambda}, \text{Res}(\hat{V}, \tilde{\Lambda}), \hat{Y}$)
10:      $\mu_1 \leftarrow \min([\Lambda \ \tilde{\Lambda}])$ ;    $\mu_{nev+nex} \leftarrow \max([\Lambda \ \tilde{\Lambda}])$
11:      $c \leftarrow \frac{b_{sup} + \mu_{nev+nex}}{2}$ ;    $e \leftarrow \frac{b_{sup} - \mu_{nev+nex}}{2}$
12:      **for** $a = 1 \rightarrow$ size($\hat{V}$) **do**
13:          $m_a \leftarrow$ DEGREES(tol, Res ($\hat{V}_{:,a}, \tilde{\lambda}_a$), $\lambda_a$, $c$, $e$)        ▷ COMPUTE POLYNOMIAL DEGREE
14:      **end for**
15:      Sort Res($\hat{V}, \tilde{\Lambda}$), $\hat{V}, \tilde{\Lambda}, m$ according to $m$
16: **end while**

---

The algorithm performs best when it receives approximate solutions which act as a pre-conditioner for the Chebyshev filter. Moreover, a rough knowledge about the eigenspectrum interval to be filtered out allows the almost immediate engagement of the approximate vectors by the filter. This knowledge is provided by inputting reasonable estimates $\mu_1$ and $\mu_{nev+nex}$ for the solutions $\lambda_1$ and $\lambda_{nev+nex}$. If these estimates, as well as the starting vectors $\hat{V}$, are not available, they can be computed by a call to a customized LANCZOS routine (see Sec. 4.2 for details).

The LANCZOS routine is the first procedure to be called within ChASE and usually it performs a handful of Lanczos steps to estimate a value $b_{sup}$ bounding $\lambda_n$ from above [52]. When $\mu_1$ and $\mu_{\text{nev+nex}}$ and/or $\hat{V}$ are not provided as input, it repeats the same number of simple Lanczos steps (i.e. a Lanczos process) multiple times with distinct randomly generated starting vectors. Averaging over all the Lanczos processes leads to the construction of an approximate spectral density [29] from which reliable estimates of $\lambda_1$ and $\lambda_{\text{nev+nex}}$ can be computed. From the last Lanczos process the routine also extracts a number of vectors that provide reasonable starting vectors. Since these are, in general, fewer than the number required by the eigensolver, they are interlaced with random vectors before being returned.

The output of the LANCZOS routine is passed to the Chebyshev FILTER routine (Alg. 1, line 4) that enhances the components of the input vectors along the directions given by the eigenvectors spanning the desired eigenspace. This is obtained by the traditional use of the three-term recurrence relation for the Chebyshev polynomials which executes a number of matrix-matrix operations between the matrix $A$ and the filtered vectors $\hat{V}$. Notice that the FILTER is called within a **while** loop which constitutes the core of SI; while the first call of the filter is executed with the same low polynomial degree ($\approx 10$) for all the vectors (Alg. 1, line 1), the second call uses an array of degrees $m$ computed on the fly and optimized to each of the vectors (Alg. 1, line 13). Details on the optimization are provided in Sec. 4.1. The block of vectors outputted by the filter spans a search space approximating a subspace containing the desired eigenspace, but the vectors can easily become linearly dependent. In order to correct such behavior the algorithm uses an orthogonalization procedure after each filtering step. In ChASE the orthogonalization is implemented through a QR factorization based on Householder reflectors (Alg. 1, line 5).

The resulting orthonormal vectors $\hat{Q}$ are then used to form a Rayleigh-Ritz quotient $G$. The quotient represents a projection of the eigenproblem $A$ onto an active subspace approximating the sought after eigenspace. The reduced eigenproblem $G$ is diagonalized by calling an external routine from a parallel library, which we generically indicate as TSOLVE (Alg. 2, line 3). In ChASE there is no preferred choice for the direct solver and depending on the library used it varies from Divide and Conquer (D&C) [11, 19] to Multiple Relative Robust Representation (MRRR) [7, 12]. The computed eigenvectors $\hat{W}$ together with the eigenvalues $\tilde{\Lambda}$ constitute the computed eigenpairs. The whole Rayleigh-Ritz procedure is described in a separate algorithm (Alg. 2) and is encoded in a separate routine in the ChASE eigensolver.

After the Rayleigh-Ritz step, the tentative pairs $(\hat{V}, \tilde{\Lambda})$ are then used to compute residuals: if the residuals are below the established threshold tol, they are accepted, locked in $\hat{Y}$ and deflated from $\hat{V}$ (see Sec. 3.4), otherwise they are used as new starting vectors in the **while** loop. For each non-converged vector, the residual and ratio of convergence are updated so that an optimized degree for the polynomial filter can be computed anew (Alg. 1, line 13). The procedure terminates when the number of consecutive eigenpairs having residuals below threshold is equal or larger than nev.

---

**Algorithm 2** Rayleigh-Ritz

1: **procedure** RAYLEIGH-RITZ($A$, $\hat{Q}$)
2:     Compute Rayleigh quotient $G = \hat{Q}^{\dagger} A \hat{Q}$          ▷ PROJECT EIGENPROBLEM TO ACTIVE SUBSPACE
3:     $(\hat{W}, \tilde{\Lambda}) \leftarrow$ TSOLVE($G$)          ▷ EITHER D&C OR MRRR
4:     **return** $(\hat{Q} \hat{W}, \tilde{\Lambda})$.
5: **end procedure**

---

### 3.3 Input, Output and Search Space

The main routine of ChASE requires a number of mandatory parameters and some optional ones. Besides the matrix $A$, standard inputs are: 1) the number of required extremal eigenpairs nev, 2) the minimum required tolerance for the eigenpair residuals tol, and 3) the generic polynomial degree deg. ChASE uses nev to set up the minimal size of the subspace which is incremented by the value of an additional required parameter nex. The combined value nev+nex denotes the total size of the search subspace. A good choice of nex represents a trade-off between extra computation and an enhanced eigenpair convergence ratio. A bigger nex increases the distance $|\lambda_{\text{nev+nex+1}} - \lambda_{\text{nev}}|$ which in turn implies higher values for $|\rho_a|$ (see Sec. 4.1), but also enlarges the size of $\hat{V}$ with a consequent increase of required FLOPs. Ideally nex should be a fraction of nev guaranteeing to include a spectral gap as large as possible between the eigenvalues $\lambda_{\text{nev}}$ and $\lambda_{\text{nev+nex}}$.

Two additional optional flags are approx and optim. The first is a flag indicating whether the user provides ChASE with information about the approximate solution of the eigenproblem (e.g. when dealing with a sequence of correlated eigenproblems) or uses ChASE in isolation as a traditional black-box solver, without any knowledge from the domain application. When the approx flag is set to `true`, ChASE expects the arrays $\hat{V}$ and $\Lambda$ to hold approximate vectors and values respectively. The smallest and largest values in $\Lambda$ are used as estimates $\mu_1$ and $\mu_{\text{nev+nex}}$ for the lower and upper end of the sought after eigenspectrum so only these two values need to be populated. On the other hand, the entire set of vectors in $\hat{V}$ are used in the Chebyshev filter as a pre-conditioner to accelerate convergence. In the output, both of these arrays are overwritten with the computed solution. If approx is set to `false`, the vectors used by the Chebyshev filter are outputted by the LANCZOS routine starting from a set of random vectors. Likewise, such a routine computes estimates $\mu_1$ and $\mu_{\text{nev+nex}}$ (see Sec. 4.2).

The optim flag specifies whether ChASE uses the same polynomial degree for all the vectors to be filtered or computes on the fly an array of optimized degrees, one for each vector of the search space. When optim is set to `false` the value stored in deg is passed to the FILTER routine each time it is called in Alg. 1. When this option is selected, we advise to use for deg a value, in double precision, larger than 20. In case ChASE is called with the optim flag set to `true`, the value stored in deg is passed to the FILTER only during the first call, while for all successive calls in the **while** loop, the FILTER routine receives an array of optimal degrees (Alg. 1, line 13).

The array of initial vectors $\hat{V}$ play a special role in SI. Already in the *ritzit* program [39] Rutishauser allows the user to provide a parameter to input [...starting values to the iteration vectors...]. He also recognizes that a poor choice of vectors orthogonal to the sought after eigenspace can negatively influence convergence. Consequently a substitution is enforced in *ritzit*: after few subspace iterations the vector with inner-most Rayleigh-Ritz projection is replaced by a random vector. Such an expedient is motivated by the following argument (see [46] p.60): since any starting vector can be represented as $\hat{v} = c_1 \hat{x}_1 + c_\perp \hat{X}_\perp$ then a measure of the distance of $\hat{v}$ from the dominant eigenvector $\hat{x}_1$ is given by the ratio $\frac{c_\perp}{c_1}$. If explicitly derived, the same ratio would appear in the expression on the right hand side of Eq. (1) for $a = 1$, determining the effective convergence of the input vector (see [46] p.57). One can also reverse the argument in order to understand the importance of using an approximate solution; having a starting vector which is already a good approximation to the desired eigenvector can substantially accelerate the solution. This specific observation is at the heart of the ChASE algorithm which targets exactly those sequences of correlated eigenproblems where the solution of the $\ell$ problem provides a good set of starting vectors for the $\ell + 1$ problem. The end result is a magnification of ChASE convergence as a function of the sequence index [14].

### 3.4 Deflation & Locking

Further filtering eigenpairs $(\hat{V}_{:,a}, \tilde{\lambda}_a)$ that satisfy the condition $\text{Res}(\hat{V}_{:,a}, \tilde{\lambda}_a) < \text{tol}$ would unnecessarily decrease the value of their residual and substantially increase the number of FLOPs performed by ChASE. For this simple reason it is good practice to remove the converged vectors and values from the arrays $\hat{V}$ and $\tilde{\Lambda}$ (Alg. 3, line 8) and store them in the respective output arrays $\hat{Y}$ and $\Lambda$ (Alg. 3, line 7). These two operations go under the conventional names of Deflation and Locking and are included in the ChASE algorithm (Alg. 1, line 9). Their overall effect is to reduce the FLOPs-count by converging the whole subspace in cumulative chunks.

While the advantage is evident, this procedure can introduce convergence issues when the tol parameter is several orders of magnitude larger than the selected machine precision. A rather high threshold tolerance for the residual may cause locking eigenpairs quite early on in the execution of ChASE. Such locked vectors may include "directions" along the sought after eigenspace which are no longer accessible by the remaining search space spanned by the vectors left in $\hat{V}$. This happens when $\hat{V}$ is re-orthogonalized against $\hat{Y}$ (Alg. 1, line 5) at the next execution of the **while** loop. This phenomenon, known as *early locking* can, in rare cases, cause the stagnation of successive eigenpairs and the consequent lack of convergence for ChASE. A failsafe mechanism for the early locking problem was proposed by Stathopoulos [42]. In the ChASE algorithm we have not introduced such a mechanism but suggest, as a rule of thumb, to avoid selecting tol $> 10^{-8}$ in double precision and tol $> 10^{-4}$ in single precision. Alternatively, the selection of a lower accuracy for the QR procedure can help to avoid stagnation in most practical cases.

It may sometimes happen—especially in the case of tight clusters of eigenvalues—that some non-consecutive eigenpairs are declared converged. This is not a problem if such eigenpairs are within the desired eigenspace of size nev. Because the size of the active subspace is larger than nev, it is not uncommon that some of these converged non-consecutive eigenpairs are outside the sought after spectrum but end up locked in the nev required eigenpairs. For this reason, as part of the deflation & locking procedure, we sort the approximate eigenpairs and their corresponding residuals according to the computed approximate eigenvalues $\tilde{\Lambda}$ (Alg. 3, line 2). Thanks to the fact that eigenvalues converge much faster than the eigenvectors (quadratically vs. linearly), in all practical cases sorting the approximate pairs guarantees that only consecutive converged eigenpairs are locked and deflated.

---

**Algorithm 3** Deflation and Locking

---

1:  **procedure** DEFLATION & LOCKING$(\hat{V}, \tilde{\Lambda}, \text{Res}(\hat{V}, \tilde{\Lambda}))$
2:      Sort $\text{Res}(\hat{V}, \tilde{\Lambda}), \hat{V}, \tilde{\Lambda}$ according to $\tilde{\Lambda}$
3:      **for** $a = 1 \rightarrow (\text{nev}+\text{nex}-\text{nconv})$ **do**
4:          **if** $\text{Res}(\hat{V}_{:,a}, \tilde{\lambda}_a) > \text{tol}$ **then**
5:              **return** $(\hat{V}, \Lambda, \hat{Y})$
6:          **end if**
7:          $\Lambda.\text{PUSH}(\tilde{\lambda}_a)$ and $\hat{Y}.\text{PUSH}(\hat{V}_{:,a})$
8:          $\hat{V}.\text{REMOVE}(\hat{V}_{:,a})$ and $\tilde{\Lambda}.\text{REMOVE}(\tilde{\lambda}_a)$
9:      **end for**
10: **end procedure**

---

## 4 ALGORITHMIC CONTRIBUTIONS

This section contains our original contributions to Algorithm 1. First, we introduce the reader to the Chebyshev filter, discuss a method to optimize the degrees of the filter polynomial $m_a$, and show how such an optimization often reduces the cost of executing the algorithm significantly. Second, we present methods to approximate $\mu_1$, $\mu_{\text{nev+nex}}$, and $b_{\text{sup}}$ which are crucial for the correct functioning of the Chebyshev filter.

## 4.1 Degree Optimization of the Chebyshev Accelerator

In the following we give a detailed introduction to the use of the Chebyshev polynomials as enhancer of the components of $\hat{V}$ that are parallel to the eigenvectors spanning the sought after eigenspace. This introduction is followed by a description of how to optimize the filtering degree, the corresponding optimizing algorithm, and its implementation in the Chebyshev filter. We conclude with some numerical results supporting our claim of increased performance.

### 4.1.1 Chebyshev polynomials and convergence ratio.

*Definition 4.1 (Chebyshev Polynomials).* A Chebyshev Polynomial $C_m(t)$ of degree $m$ is defined on the real axis as

$$C_m(t) = \cosh\left(m \cosh^{-1}(t)\right), \quad t \in \mathbb{R} \tag{2}$$

When constrained on the interval $[-1, 1]$ such definition naturally reduces to $C_m(t) = \cos\left(m \cos^{-1}(t)\right)$.

While on the interval $[-1, 1]$ the Chebyshev polynomial of degree $m$ is an oscillating function with $m - 1$ extrema, for $|t| > 1$ the function diverges quite rapidly already for modest values of $m$. The polynomials $C_m(t)$ can be also defined through a three-terms recurrence relation

$$C_{m+1}(t) = 2t C_m(t) - C_{m-1}(t), \tag{3}$$

where the first two polynomials are equal to $C_0(t) = 1$ and $C_1(t) = t$. The asymptotic behavior of the polynomials for $|t| > 1$ can be quantified by expressing $C_m(t)$ as an explicit function of $t$. By defining

$$y \doteq \cosh^{-1}(t) \quad \text{and} \quad \rho \doteq \exp(y)$$

it is straightforward to show that $t$ and $\rho$ are related through a quadratic equation

$$\rho^2 - 2t\rho + 1 = 0$$

admitting $\rho$ and $\rho^{-1}$ as solutions. By convention we choose for a fixed $t$

$$|\rho| = \max\left|t \pm \sqrt{t^2 - 1}\right| \quad \text{and} \quad |\rho|^{-1} = \min\left|t \pm \sqrt{t^2 - 1}\right|$$

and plug them back in the definition of $y$ so that the polynomial can be re-written as

$$C_m(t) = \cosh\left(m \ln|\rho|\right) = \frac{|\rho|^m + |\rho|^{-m}}{2}. \tag{4}$$

For $|t| > 1$, $|\rho|$ is always larger than one and the leading asymptotic behavior of the polynomial for increasing $|t|$ is given by $C_m(t) \sim \frac{|\rho|^m}{2}$. In other words, outside of the interval $[-1, 1]$ $C_m(t)$ diverges as a polynomial of degree $m$. The Chebyshev filter is based on a minimax theorem (see [40, p. 109] for a proof) from function approximation theory

THEOREM 4.2. *Let $|s| > 1$ and $\mathbb{P}_m$ denote the set of polynomials of degree smaller or equal to $m$. Then the extremum*

$$\min_{p \in \mathbb{P}_m, p(s)=1} \max_{t \in [-1, 1]} |p(t)|$$

*is reached by*

$$p_m(t) \doteq \frac{C_m(t)}{C_m(s)}.$$

This theorem ensures that $p_m(t)$ is the smallest polynomial of degree $m$ inside the interval $[-1, 1]$ for any chosen point $s$ outside of it. In the limit of large degrees such polynomials have an asymptotic limit

$$p_m(t) = \frac{C_m(t)}{C_m(s)} \xrightarrow{\text{large } m} \frac{1}{|\rho_s|^m} \quad \text{with} \quad |\rho_s| = \max_{\{\pm:\, |s|>1\}} \left| s \pm \sqrt{s^2 - 1} \right|,$$

where only the leading term of Eq. (4) is kept together with $\sup_{t \in [-1,1]} C_m(t) = 1$. For a generic interval $[\alpha, \beta] \subset \mathbb{R}$, one defines the center of the interval $c = \frac{\alpha+\beta}{2}$ and the half-width of the interval $e = \frac{\beta-\alpha}{2}$, and builds a linear transformation $x(t) = c + e \cdot t$ mapping the interval $[-1, 1]$ onto $[\alpha, \beta]$. Using this transformation for $t$ and $s$, we obtain Chebyshev polynomials for a general interval

$$p_m(x) \doteq \frac{C_m(\frac{x-c}{e})}{C_m(\frac{\gamma-c}{e})} \xrightarrow{\text{large } m} \frac{1}{|\rho_\gamma|^m} \quad \text{with} \quad |\rho_\gamma| = \max_{\{\pm:\, \gamma \notin [\alpha, \beta]\}} \left| \frac{\gamma - c}{e} \pm \sqrt{\left(\frac{\gamma - c}{e}\right)^2 - 1} \right|. \quad (5)$$

Theorem 4.2, and recent numerical analysis results [13], suggest that the polynomial $p_m(t)$ can be exploited for enhancing the components of the vectors $\hat{V}$ along the eigenvectors corresponding to the sought after spectrum. In practice, by equating $\mu_1 = \gamma$, $\mu_{\text{nev+nex}} = \alpha$, and $b_{\text{sup}} = \beta$, one can show [13] that

$$\min_{v \in \mathcal{V}} \|\hat{v} - \hat{x}_a\| \leq \eta_a |\rho_a|^{-m}, \quad (6)$$

where $\mathcal{V} = \text{span}\left(p_m(A) \cdot \hat{V}\right)$ and $\eta_a$ is a constant. Thus, the subspace spanned by the filtered vectors always contains a vector which converges to the eigenvector $x_a$, corresponding to an eigenvalue $\lambda_a \in [\mu_1, \mu_{\text{nev+nex}}]$, with a convergence ratio equal to $|\rho_a|^{-1}$. The result just stated makes it possible to use an opportunely adapted 3-terms recurrence relation [40] that defines the action of the $p_m$ polynomial onto the $\hat{V}$ vectors

$$p_{m+1}(\hat{V}) = 2 \frac{\sigma_{m+1}}{e}(A - cI)\, p_m(\hat{V}) - \sigma_{m+1}\sigma_m\, p_{m-1}(\hat{V}), \quad (7)$$

with $p_m(\hat{V}) = p_m(A) \cdot \hat{V}$, $\sigma_m = \left(\frac{2}{\sigma_1} - \sigma_{m-1}\right)^{-1}$, and $\sigma_1 = \left(\frac{\gamma-c}{e}\right)^{-1}$. The relation above is algorithmically encoded in the non-colored part of the procedure in Algorithm 4. The output has the same number of vectors as the input, which are progressively aligned with the active search space, and have components orthogonal to such space converging to zero with ratios proportional to the ones reported in Eq. (6).

*4.1.2 Derivation of the minimal degree.* Conventional polynomial filtering uses a fixed degree of the Chebyshev filter, as in the unmodified version of Alg. 1. Even when the degree is a configuration parameter specified by the user, two problems remain: first, eigenvectors close to convergence are filtered more than necessary, resulting in unnecessary work. On the other hand, eigenvectors far from convergence are not filtered enough, requiring more iterations of the **while** loop in Alg. 1. Choosing the degree of the filter becomes a trade-off between the number of required iterations and excessive filtering of nearly converged vectors. This trade-off can be avoided entirely by *optimizing the degree* of the filter separately for each vector in Alg. 4. This section briefly discusses the derivation of the minimal polynomial degree for each vector, and the necessary algorithmic changes. We will see that, thanks to the degree optimization, ChASE can save up to 20% of required FLOPs.

The result reported in Eq. (6) suggests that eigenvectors having residuals with a given level of tolerance can be obtained using a Chebyshev filter with a polynomial of minimal degree $m_{\text{min}}$[2] for

---

[2]Mention of a tailored polynomial degree was already suggested in [38] and [46, p.395]. A rigorous analysis is provided in [13].

---

**Algorithm 4** Chebyshev filter for optimized even degrees

---

1: **procedure** FILTER($A$, $b_{sup}$, $\mu_1$, $\mu_{nev+nex}$, $\hat{V}$, sorted $m = [m_1 \dots m_{size(\hat{V})}]$)
2:      $c = \frac{b_{sup}+\mu_{nev+nex}}{2}$ ;    $e = \frac{b_{sup}-\mu_{nev+nex}}{2}$
3:      $\sigma_1 = \frac{e}{\mu_1-c}$
4:      $\hat{W} \leftarrow \frac{\sigma_1}{e}(A - c\,I_n)\,\hat{V}$
5:      $\sigma \leftarrow \sigma_1$
6:      $s \leftarrow 1$
7:      **for** $i = 2, \dots, m_{size(\hat{V})}$ **do**
8:          $\tau \leftarrow \frac{1}{\frac{2}{\sigma_1}-\sigma}$
9:          $\hat{V}_{:,s:end} \leftarrow 2\,\frac{\tau}{e}(A - c\,I_n)\,\hat{W}_{:,s:end} - \tau\,\sigma\,\hat{V}_{:,s:end}$
10:         $\hat{V} \longleftrightarrow \hat{W}$
11:         $\sigma \leftarrow \tau$
12:         **while** $m_s \leq i$ **do**
13:             $s \leftarrow s + 1$
14:         **end while**
15:      **end for**
16:      **return** $\hat{W}$
17: **end procedure**

---

each of the vectors $\hat{V}_{:,a}$. This minimal value for $m$ can be computed using a simple strategy. One executes the very first instance of the **while** loop in Alg. 1 using an initial low degree $m_0$ (between 8 and 15) for filtering all the $\hat{V}$ (Alg. 1, line 1). At the end of the first loop, all residuals for the approximate eigenpairs are computed Res($\hat{V}_{:,a}^{(0)}, \tilde{\lambda}_a$) (Alg. 1, line 8), where for sake of conciseness we indicate $\hat{V}_{:,a}^{(i)} = p_{m^{(i)}}(A) \cdot \hat{V}_{:,a}$. Since the curve describing the residuals of the approximate solutions as a function of $m$ is a function of the convergence ratio for each single eigenvector [13], the next filtering step within the **while** loop will produce a residual equal to

$$\text{Res}(\hat{V}_{:,a}^{(1)}, \tilde{\lambda}_a) = \frac{\text{Res}(\hat{V}_{:,a}^{(0)}, \tilde{\lambda}_a)}{|\rho_a|^{m_a^{(1)}}}.$$

The requirement that Res($\hat{V}_{:,a}^{(1)}, \tilde{\lambda}_a$) $\leq$ tol immediately translates into the condition

$$m_a^{(1)} \geq \frac{\log\left|\frac{\text{Res}(\hat{V}_{:,a}^{(0)}, \tilde{\lambda}_a)}{\text{tol}}\right|}{\log|\rho_a|}. \tag{8}$$

Because $m_{min}$ depends on $|\rho_a|$, to each approximate eigenvector corresponds a different minimal degree value $m_a$. By using a filter with a polynomial degree tailored to the specific approximate eigenpair, one ensures that some eigenpairs may already converge at the second filtering step. Moreover the filtering step will perform the minimal amount of matrix-vector operations necessary for the eigenpair to converge. In other words, the computational cost of the filtering step has been optimized.

*4.1.3 Implementation notes.* The procedure for obtaining the optimal degree for a single approximate eigenpair is outlined in Alg. 5. In addition to the center $c$ and half-width $e$ of the interval $[\mu_{nev+nex}, b_{sup}]$, the algorithm requires the tolerance threshold tol below which eigenpairs are declared converged, the eigenvalue $\tilde{\lambda}_a$, and the residual for the unconverged eigenpair Res($\hat{V}_{:,a}, \tilde{\lambda}_a$).

---

**Algorithm 5** Degree Optimization

---

1: **procedure** DEGREES(tol, Res($\hat{V}_{:,a}, \tilde{\lambda}_a$), $\tilde{\lambda}_a$, $c$, $e$ )
2:      $t = \frac{\tilde{\lambda}_a - c}{e}$
3:      $|\rho| = \max \left\{ \left| t - \sqrt{t^2 - 1} \right|, \left| t + \sqrt{t^2 - 1} \right| \right\}$
4:      deg $= \left\lceil \left| \log \frac{\text{Res}(\hat{V}_{:,a}, \tilde{\lambda}_a)}{\text{tol}} \, / \, \log |\rho| \right| \right\rceil$               ▷ DEGREE REQUIRED FOR CONVERGENCE
5:      deg $= \min \{$deg $+$ DEGEXTRA, DEGMAX$\}$                  ▷ DEGEXTRA $\approx 2$
6:      **return** deg $+$ mod(deg, 2)                            ▷ ENSURE deg IS EVEN
7: **end procedure**

---

The degree estimate, computed using the lower value satisfying the inequality in Eq. (8), is incremented by a small amount, DEGEXTRA $\approx 2$, to account for rounding errors in the degree computation. In addition, we introduce an upper threshold DEGMAX for the largest admissible value of the minimal degree $m_a$, as too large of a degree causes the filtered subspace $\hat{V}$ to become rank deficient. In such cases orthogonalization via a QR factorization introduces numerical instabilities, resulting in a systematic increase of the residuals of the non-converged vectors [13]. In our experience a maximum degree of 36 (for double precision) effectively prevents the appearance of divergent residuals (Alg. 5, `line` 5).

Including the optimization of the filter degree in the ChASE algorithm requires some changes and additions, which are color coded in blue both in Alg. 1 and 5. Degree optimization calls for knowledge of the residuals (Alg. 1, `line` 8), and as such is only possible from the second iteration of the **while** loop onwards. Hence, the initial filtering is carried out with a low polynomial degree (Alg. 1, `line` 1). A high performance implementation of the Chebyshev filter involves additional changes to Alg. 4. In the matrix-matrix multiplication (expressed as calls to the BLAS-3 HEMM kernel) in `line` 9 and `line` 4, each vector of $\hat{V}$ is filtered with an optimized and generally distinct polynomial degree $m_a$. Consequently, the multiplication has to be implemented such as to omit, from $\hat{V}$ and $\hat{W}$, vectors already filtered up to the desired degree. To maximize performance, it is preferable to keep a single HEMM call regardless of which vectors have to be omitted. Sorting $\hat{V}$ according to the polynomial degree (Alg. 1, `line` 15) forces all matrix-matrix multiplications to be on contiguous vectors and thus can be achieved with a single call to the HEMM routine. Each time the smallest entry $m_s$ in the array of degrees becomes larger than the iterator $i$ of the **for** loop (Alg. 4, `line` 7), the relative vectors are removed from the next HEMM call and so on and so forth, until the iterator equals $m_{\text{size}(\hat{V})}$ (Alg. 4, `lines` 12 -- 14).

We conclude with a minor but important remark: the double buffering in line 9 of Alg. 5 (used to implement Eq. (7)) causes any vector filtered with an odd degree to be saved to $\hat{V}$. However, since the FILTER routine returns the $\hat{W}$ array, any such vector would need to be copied from $\hat{V}$ to $\hat{W}$. We obviate the need for data movement by constraining the outputted array of degrees deg to contain only even values (Alg. 5, `line` 6).

*4.1.4 Optimization and performance.* Given the residuals Res($\hat{V}_{:,a}, \tilde{\Lambda}_a$) for each eigenpair, the calculation of the array of optimal degrees is quite inexpensive (see Alg. 5), and has an overall positive impact on ChASE's performance. Minimizing the FLOPs executed by the FILTER routine accounts for a reduced time to solution. How much the minimization influences ChASE's performance depends on two distinct factors: the choice of the initial polynomial degree—which, in the absence of degree optimization, is the polynomial degree used in each call of the FILTER procedure for all vectors $\hat{V}$—and the context in which ChASE is employed. For the latter, we show that the impact on its performance is maximized when ChASE is used to tackle sequences of eigenvalue problems.

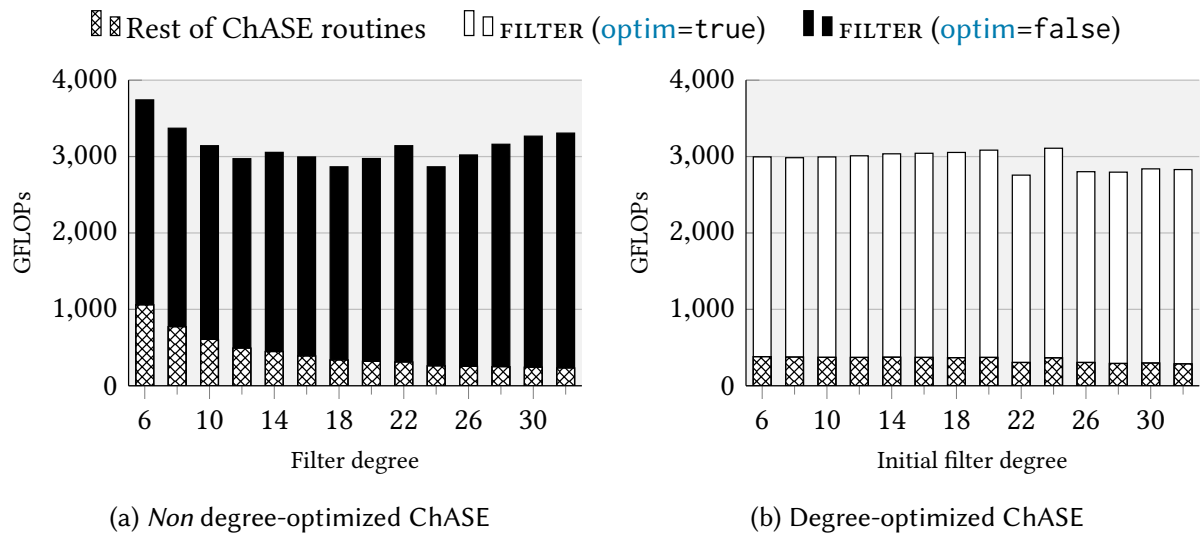

(a) *Non* degree-optimized ChASE     (b) Degree-optimized ChASE

Fig. 1. GFLOPs for ChASE with (optim=true) and without (optim=false) degree optimization.

For the experiments in this section we use a sequence of eigenproblems obtained from a DFT simulation using the FLEUR code [9], and labeled NaCl. The matrix size of the eigenvalue problems is $n = 3893$ and we seek the smallest nev = 256 eigenpairs, corresponding to about 7% of the spectrum. For the ChASE tol and nex parameters we selected the values of $10^{-10}$ and 26, respectively. The presence of a spectral gap between the first nev and the next nev+nex eigenvalues causes ChASE to converge quite well for each problem in the sequence.

An appropriate choice of the initial/constant degree is particularly important in order to obtain high performance. As discussed at the beginning of the section, choosing a large degree results in "over-filtering" eigenvectors that are already converged. Conversely, a small polynomial degree avoids over-filtering at the cost of additional iterations. This trade-off is visualized in Figure 1a, which plots the (constant) filter degree against the GigaFLOPs (GFLOPs) required by ChASE to converge. These results are obtained with the first problem in the NaCl sequence and a random i.i.d. collection of vectors $\hat{V}$ as input. The GFLOPs not spent in the Chebyshev filter are indicated as crosshatched in the bar plot; this includes, for example, the Rayleigh-Ritz procedure, the QR factorization, and the computation of the residuals. For low polynomial degrees more GFLOPs are spent outside of the filter, indicating a larger number of iterations of the **while** loop in Alg. 1. The total number of FLOPs forms a rough "bathtub" shape, with the optimum filter degree between 18 and 24: a high polynomial degree causes over-filtering and increases the total number of required GFLOPs.

Figure 1b shows the same experiment setup as Figure 1a, but with an optimized array of degrees $m$. Here the degree specified on the x-axis is the initial degree, i.e. the polynomial degree deg used in the first ChASE iteration. As a result, the number of GFLOPs required for ChASE to converge is practically independent of the initial degree. The degree optimization acts as a sort of stabilizer of the FLOPs count against variations of deg: the choice of the initial degree has only a minimal impact on the eigensolver's performance. As expected, ChASE needs fewer GFLOPs to converge for the majority of the deg values, implying that degree optimization yields, in most cases, a faster solver.

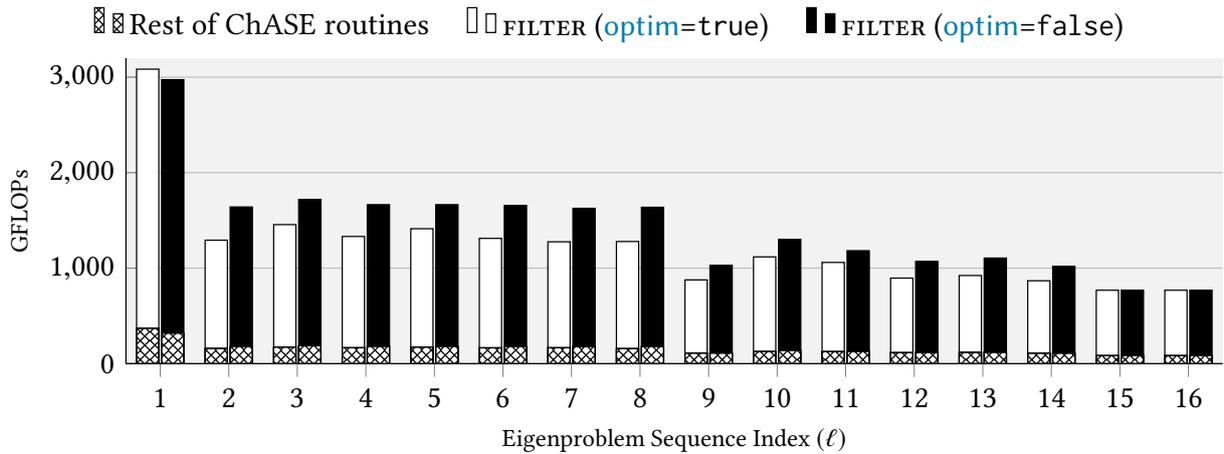

Fig. 2. Optimized (optim=true) vs. non-optimized (optim=false) FILTER for the NaCl ($n$=3,893, nev=256) sequence

On sequences of eigenvalue problems the impact of degree optimization is even more dramatic. Figure 2 compares the GFLOPs required for ChASE to converge with, and without, degree optimization. The data in the plot refers to the NaCl sequence, where the first problem $P^{(1)}$ is started with a random i.i.d. set of vectors $\hat{V}$. For the rest of the sequence, the problem $P^{(\ell)}$ is solved by inputting to ChASE the solution $\hat{Y}^{(\ell-1)}$ of the problem $P^{(\ell-1)}$. We select for deg an initial value of 20, a choice that favors the use of ChASE with a constant degree. Such a value—a worst-case scenario for degree optimization—shows up in the first problem of the sequence where ChASE with degree optimization performs slightly worse than ChASE with constant degree. For most of the remaining problems, the degree optimization saves up to 25% of the required FLOPs as compared to a constant degree. Notable exceptions are the last two problems with $\ell = 15, 16$; these problems converge within a single iteration and thus there is no practical difference in the execution of ChASE with optimized or with constant degrees. As in the previous figure, the crosshatched area indicates GFLOPs spent in ChASE routines other than the polynomial filter: for this area there is very little difference between the two versions of ChASE, as the degree optimization does not directly influence the number of **while** loop iterations in this sequence. A more in-depth analysis of ChASE's performance for sequences is provided in Section 6.

## 4.2 Spectral Estimates

The spectral estimates—$\mu_1$, $\mu_{\text{nev+nex}}$, and $b_{\text{sup}}$—are of critical importance for both correctness and performance of ChASE. The estimates are used in two separate but related procedures: the Chebyshev filter (Alg. 4) and the degree optimization (Alg. 5). Given the value of $\mu_1$, the filter is constructed to suppress the spectrum between $\mu_{\text{nev+nex}}$ and $b_{\text{sup}}$. In the degree optimization the spectral estimates are used to compute the ratios of convergence $|\rho_a|^{-1}$. In this section we illustrate how to obtain $\mu_1$, $\mu_{\text{nev+nex}}$, and $b_{\text{sup}}$, both with and without the availability of an approximate solution. We discuss the performance implications in the choice of the bounds, as well as the cost of obtaining accurate estimates.

$\mu_1$ and $\mu_{\text{nev+nex}}$ are approximations of the first and (nev+nex)$^{\text{th}}$ eigenvalue, respectively. After the first iteration of the **while** loop, new approximations for these values can be obtained from the Ritz values computed in Alg. 2. Even when initial values for $\mu_1$ and $\mu_{\text{nev+nex}}$ are inaccurate, ChASE capitalizes on the quadratic convergence of the Ritz values and self-corrects $\mu_1$ and $\mu_{\text{nev+nex}}$ after only a few iterations. In spite of its robustness, this self-correcting behavior is sometimes detrimental

---

**Algorithm 6** Lanczos procedure for parameters estimates

---

1: **procedure** LANCZOS($A$, approx)
2:     $AU = UT_k + f_k e_k^\top$     $T_k = Z^H \tilde{\Lambda}_k Z$     $\tilde{\Lambda}_k = \text{diag}[\tilde{\lambda}_1, \ldots, \tilde{\lambda}_k]$     ▷ $k$ Lanczos steps
3:     $b_{\text{sup}} = \|f_k\|_2 + \max[\tilde{\lambda}_1, \ldots, \tilde{\lambda}_k]$
4:     **if** approx **then**
5:         **return** $b_{\text{sup}}$
6:     **end if**
7:     **for** $j = 1, \ldots, n_{\text{vec}}$ **do**
8:         $AU^{[j]} = U^{[j]} T_k^{[j]} + f_k^{[j]} e_k^\top$     $T_k^{[j]} = (Z^{[j]})^H \tilde{\Lambda}_k^{[j]} Z^{[j]}$     ▷ $k$ Lanczos steps
9:     **end for**
10:    Construct DoS: $\tilde{\phi}(t) = \frac{1}{n_{\text{vec}}} \sum_{j=1}^{n_{\text{vec}}} \sum_{i=0}^{k} (Z_{1,i}^{[j]})^2 g_\sigma(t - \tilde{\lambda}_i^{[j]})$     ▷ See [29, Section 3.2]
11:    Find $\tilde{t}$ such that $\int_{-\infty}^{\tilde{t}} \tilde{\phi}(t) dt \approx \frac{\text{nev+nex}}{N}$
12:    $\mu_{\text{nev+nex}} = \tilde{t}$     $\mu_1 = \min_{j=1} \tilde{\lambda}_1^{[j]}$
13:    Intersperse $Z^{[1]} Q^{[1]}$ corresponding to $\tilde{\lambda}^{[1]} < \mu_{\text{nev+nex}}$ into $\hat{V}$
14:    **return** $\mu_1$, $\mu_{\text{nev+nex}}$, $b_{\text{sup}}$, and $\hat{V}$.
15: **end procedure**

---

of the overall algorithm efficiency. For instance, over-correcting the estimate for $\mu_{\text{nev+nex}}$ makes the filter less effective, and similarly the degree optimization does not produce accurate results. Contrary to $\mu_1$ and $\mu_{\text{nev+nex}}$, $b_{\text{sup}}$ is always computed at the beginning of the ChASE algorithm and must be an accurate upper bound of the largest eigenvalue, otherwise ChASE will fail because some of the largest eigenvalues will not be filtered out. For all the reasons above, we wish to obtain an approximation of $\mu_1$, $\mu_{\text{nev+nex}}$, and $b_{\text{sup}}$ that is at the same time accurate and inexpensive.

In the case of sequences of correlated eigenvalue problems, $\mu_1$ and $\mu_{\text{nev+nex}}$ need to be approximated only for the first problem of the sequence. Successive problems can use the approximate spectrum of $P^{(\ell-1)}$ to obtain the estimates for $P^{(\ell)}$. However, in order to ensure that $b_{\text{sup}}$ is an upper bound of the spectrum it must be recomputed for each new problem $P^{(\ell)}$. Estimates for $b_{\text{sup}}$ with a varying degree of accuracy can be inexpensively computed using the method in [52]. Estimating an interior eigenvalue is quite more complicated and expensive than estimating the extrema of the spectrum. To obtain an estimate for $\mu_{\text{nev+nex}}$, which is usually in the first quarter of the spectrum, we propose to employ an approximation of the spectral density, sometimes referred to as Density of States (DoS) [29, Section 3.2.1]. The DoS function can be interpreted as a spectral probability density function. In its regularized form it is given by

$$\phi(t) = \frac{1}{n} \sum_{j=1}^{n} g_\sigma(t - \lambda_j), \tag{9}$$

where $g_\sigma$ is the Gaussian kernel of standard deviation $\sigma$, and controls the smoothness of the resulting function.

Naturally, we want to approximate $\phi$ without requiring all eigenvalues of the matrix. To this end, let $AU = UT_k + f_k e_k^\top$ be the result of $k$ steps of the Lanczos method with a starting vector $v_0$. Let's further write the tridiagonal symmetric matrix $T_k = Z^H \tilde{\Lambda}_k Z$, with $\tilde{\Lambda}_k = \text{diag}[\tilde{\lambda}_1, \ldots, \tilde{\lambda}_k]$. It can be shown that $\sum_{j=1}^{k} (Z_{1,j})^2 g_\sigma(t - \tilde{\lambda}_j)$ approximates a weighted DoS, where the weighting is related to the expansion coefficients of $v_0$ in the basis of $A$. By repeating the Lanczos method $n_{\text{vec}}$ times with different random vectors $v_0$ and taking the average of them, we reduce the effect of the weighting

$$\tilde{\phi}(t) = \frac{1}{n_{\text{vec}}} \sum_{i=1}^{n_{\text{vec}}} \sum_{j=1}^{k} (Z_{1,j}^{[i]})^2 g_\sigma(t - \tilde{\lambda}_j^{[i]}). \tag{10}$$

Our approximation of $\mu_{\text{nev+nex}}$ is a value $\bar{t}$, such that $\int_{-\infty}^{\bar{t}} \tilde{\phi}(t)dt \approx \frac{\text{nev+nex}}{n}$. Algorithm 6 illustrates the complete procedure. $b_{\text{sup}}$ is computed via an initial $k$ Lanczos steps and the formula given in [52, Eq. (2.5)] (Alg. 6, `line 3`). If the other two spectral approximations are required, we perform an additional $n_{\text{vec}}$ Lanczos procedures and form the DoS from Equation (10) (Alg. 6, `line 10`). Finally, we obtain $\mu_{\text{nev+nex}}$ by integrating over the spectral density. For the sake of simplicity, we obtain $\mu_1$ not via the DoS, but rather by using the smallest Ritz value of the Lanczos process. Since $ZQ$ are the approximate eigenvectors of $A$, we use those vectors corresponding to the Ritz values smaller than $\mu_{\text{nev+nex}}$ and intersperse them with random i.i.d. vectors in $\hat{V}$.

The Alg. 6 has three parameters—the number of steps $k$ for each Lanczos procedure, the number of different starting vectors $n_{\text{vec}}$, and the regularization parameter $\sigma$ for the DoS—which regulate a trade-off between accuracy and computational cost. When $\mu_{\text{nev+nex}}$ is required the LANCZOS subroutine may account for about 10% of ChASE total execution time. Nevertheless, less accurate spectral estimates degrade the effectiveness of the filter by far more than 10%. A good parameter choice must trade-off a better convergence rate of the SI against the additional time spent in the LANCZOS routine. This is complicated by the fact that the relationship between the quality of the bounds and the subspace convergence is not straightforward. For small $k$, the computational cost is dominated by the $k \, n_{\text{vec}}$ matrix-vector multiplications inside the Lanczos method. The cost for solving the tridiagonal eigenproblem, as well as the construction of $\tilde{\phi}$ and its integration, are negligible. Larger $k$ would have a number of advantages, namely better accuracy of the DoS inside of the spectrum and more approximate eigenvectors $ZQ$ (Alg. 6, `line 13`). However, in most cases the increased orthogonalization cost of large $k$ does not pay off in terms of better convergence of the SI. In double precision, we suggest the default value $k$=25. The work by Lin, Saad, and Yang [29] uses values between 40 and 100, however their goal is an accurate approximation of the entire spectral density. Since the Lanczos method tends to approximate the extrema of the spectrum rather quickly, a smaller choice of $k$ is quite reasonable to estimate $\mu_{\text{nev+nex}}$.

The regularization parameter $\sigma$ controls the width of the Gaussians $g_\sigma$ and thus the smoothness of the spectral density. Since $k$ essentially determines the number of Gaussians, there is a strong connection between the two parameters. We propose a standard value of $\sigma = 0.25$, a value close to the one used in [29]. A too small value for $n_{\text{vec}}$ results in a poor average and thus a distorted spectral density. Conversely, too large of a $n_{\text{vec}}$ is needlessly expensive without a noticeable benefit for ChASE. We recommend a value between 4 and 10 for $n_{\text{vec}}$, with the default set to 4. In concluding, we point out that the $n_{\text{vec}}$ matrix-vector multiplications are coalesced into a single HEMM, significantly increasing performance. Fine tuning the method we described in this section took considerable effort. We omit the complicated and tedious details due to space constraints. In addition the ChASE code implements a number of additional heuristics that result in values for $\mu_{\text{nev+nex}}$ that work well on the vast majority of cases. To sum up, our implementation performs almost always better than a naive choice of $\mu_{\text{nev+nex}}$.

## 5 USING CHASE

The linear algebra routines within ChASE, such as matrix-matrix multiplications and orthogonalization, are disentangled from the algorithm proper. Section 5.1, explains the underlying modularity concept. This separation of concerns allows for different parallelization approaches. In Section 5.2 we present two such approaches, ChASE-MPI and ChASE-Elemental, that come with the ChASE library. Section 5.3 introduces the configuration parameters of ChASE, as well as a code snippet that illustrates ChASE's usage in application codes[3].

## 5.1 Decoupling kernels from the ChASE algorithm

ChASE relies on a modest number of numerical kernels whose calls execute almost all floating-point operations.[4] Because of its simple structure there would be clear advantages in the separation of the numerical kernels from the algorithm implementation. Reverse Communication Interfaces (RCI) [17] are a time-honored method to decouple some operation(s), like the matrix-vector multiplication, from a given algorithm. A popular and established example is the ARPACK library [27], but RCIs are also used in more recent software packages like FEAST [34]. RCIs tend to produce complicated code because information must be passed via return values. There are modern alternatives to RCIs. In ChASE the linear algebra kernels are separated from the main algorithm through the use of an object-oriented software interface. This approach is similar to the one realized in Anasazi [2], although our implementation is simpler and leads to a number of immediate benefits.

**Listing 1** Class interface that abstracts the ChASE algorithm from the numerical kernels

```
1  using T = std::complex<double>;
2  class Chase {
3   public:
4    virtual void Start() = 0;                        // Alg. 1 line 1
5    virtual void End() = 0;                          // Alg. 1 line 16
6    virtual void Resd(double *ritzv, double *resd,   // Alg. 1 line 8
7                      size_t fixednev) = 0;
8    virtual void Lock(size_t new_converged) = 0;     // Alg. 3 line 7
9    virtual void QR(size_t fixednev) = 0;            // Alg. 1 line 5
10   virtual void RR(double *ritzv, size_t block) = 0; // Alg. 2
11   virtual void HEMM(size_t nev, T alpha, T beta, size_t s) = 0;  // Alg. 4 line 9
12   virtual void Lanczos(size_t k, double *upperb) = 0;  // Alg. 6 line 3
13   virtual void Lanczos(size_t M, size_t j, double *upperb,  // Alg. 6 line 8
14                        double *ritzv, double *Tau,
15                        double *ritzV) = 0;
16   virtual void LanczosDos(size_t idx, size_t m, T *ritzVc) = 0;  // Alg. 6 line 13
17   virtual void Shift(T c, bool isunshift = false) = 0;  // A - cI_n
18   virtual void Swap(size_t i, size_t j) = 0;       // Swap V̂_{:,i} and V̂_{:,j}
19
20   /* ommited Getters */
21  };
```

First, ChASE can be easily integrated into existing codes thanks to the relative simplicity of the software interface. For instance, the low-level kernels can be implemented according to an existing distribution of the matrix elements of $A$ so as to avoid the need to re-distribute data. Second, ChASE can easily exploit existing linear algebra libraries such as BLAS and LAPACK all the way up to GPU-based kernels, and even complex distributed-memory dense linear algebra frameworks such as Elemental [35]. Indeed, the ChASE library includes a version supporting MPI+GPUs (ChASE-MPI), as well as one that uses Elemental for the numerical kernels (ChASE-Elemental).

As a C++ program, the decoupling between the ChASE algorithm and the implementation of the numerical linear algebra kernels is accomplished via a *pure abstract class* which defines the interface for the C++ numerical kernels. A slightly simplified version of the interface is given in Listing 1. The actual implementation is templated to allow for real-valued matrices and single precision data types. A derived class of `Chase` must implement these listed kernels, including an Hermitian matrix-matrix multiply, or `HEMM`. Other kernels are more complex, such as the $k$-step Lanczos and the Rayleigh-Ritz procedure. All parallelism and handling of data is performed by classes derived from `Chase`. In addition to the advantages already mentioned, the abstraction allows the ChASE algorithm itself to be short and easily readable, with the corresponding C++ source code being very similar to Algorithm 1. As an illustrative example, we illustrate a derived class of `Chase` based on the BLAS and LAPACK kernels.

---

[3]To align our notation with the ChASE code snippets, we refer to the rank of matrix $A$ with the letter N in this section only.
[4]The only notable exception is the calculation of the Chebyshev coefficients and the polynomial degree.

*5.1.1  A sketch of ChASE with BLAS+LAPACK.* `ChaseBLAS` is an implementation of the interface given in Listing 1 using BLAS and LAPACK kernels. The Level 3 Basic Linear Algebra Subprograms (BLAS) [16] and the Linear Algebra PACKage [1] are staples of (single-node) dense linear algebra. `ChaseBLAS` is a relatively simple example and yet it is rich enough to highlight the power of the interface-based approach used by ChASE; in fact, it constitutes a high performance implementation for shared-memory computers. Listing 2 shows only a partial view of the `ChaseBLAS` class where, for

---

**Listing 2** ChaseBLAS: an implementation of ChASE's kernels with BLAS+LAPACK

```
1   using T = complex<double>;
2   class ChaseBLAS : public chase::Chase {
3    public:
4     ChaseBLAS(size_t N, size_t nev, size_t nex, T *H, T *V, T *W, double *ritzv,
5               double *resid)
6         : N_(N), nev_(nev), nex_(nex), locked_(0), H_(H), approxV_(V),
7           workspace_(W), ritzv_(ritzv), resid_(resid), config_(N_, nev_, nex_) {}
8
9     void HEMM(size_t block, T alpha, T beta, size_t s) override {
10        zhemm('L', 'L', N_, block, &alpha, H_, N_, approxV_ + N_ * (locked_ + s),
11              N_, &beta, workspace_ + N_ * (locked_ + s), N_);
12        swap(approxV_, workspace_);
13    };
14     void Lock(size_t new_converged) override {
15        memcpy(workspace_ + locked_ * N_, approxV_ + locked_ * N_,
16               N_ * (new_converged) * sizeof(T));
17        locked_ += new_converged;
18    };
19     /* Additional functions from Listing 1 ommited for conciseness */
20    private:
21     int N_, nev_, nex_, locked_;
22     T *A_, *approxV_, *workspace_;
23     double *ritzv_, resid_;
24     ChaseConfig<T> config_;
25  };
```

---

the sake of brevity, only the constructor, data members, the `HEMM`, and `Lock` functions are displayed. `ChaseBLAS` is constructed with the necessary sizes `N`, `nev`, and `nex`, as well as the required buffers for $A$, $\hat{V}$, etc. The heart of any `Chase` implementation is the `HEMM` function (Lst. 2, `line 9`), which is typically implemented as a wrapper to an optimized BLAS library (e.g., MKL, cuBLAS).

The three-terms recurrence relation of the Chebyshev filter is implemented via the `HEMM` function. `approxV_` is a column major array that contains the approximate eigenvectors sorted according

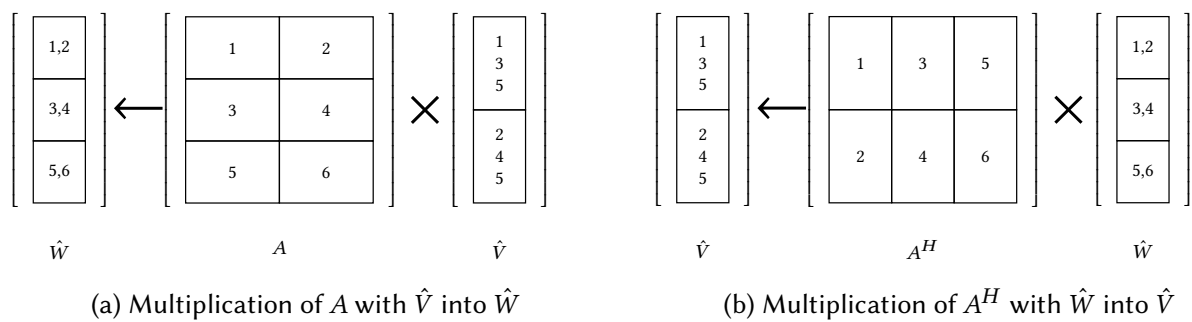

(a) Multiplication of $A$ with $\hat{V}$ into $\hat{W}$      (b) Multiplication of $A^H$ with $\hat{W}$ into $\hat{V}$

Fig. 3. Example data distribution for 6 MPI ranks. Numbers inside blocks indicate owning MPI processes. Alternating multiplications of $\hat{V}$ and $\hat{W}$ with $A$ do not require redistribution of data.

to the increasing entries of the array of degrees $m$ (see Sec. 4.1.3). A natural way to leverage the recurrence relation is via double-buffering, implemented here by swapping the pointers for `approxV_` and `workspace_`. The HEMM function is called hundreds of times during a typical solve, and performs around 80% of the total FLOPs. There are two reasons why the first `locked_`+s vectors stored in `approxV_` are excluded from the multiplication, and so they are not filtered any further. First, because they are locked vectors; in the language of Alg. 1 such vectors are not part of $\hat{V}$, but rather part of $\hat{Y}$. In the C++ implementation, ChASE does not maintain a separate $\hat{Y}$ buffer. More precisely, it moves converged vectors to the beginning of the `approxV_` buffer and locks them by calling the `Lock` function. Within `ChaseBLAS` the `locked_` variable indicates how many vectors at the beginning of the `approxV_` buffer have been locked (Lst. 2, `line 17`). Secondly, when optim=`true`, vectors may be excluded from the multiplication with $A$ if they have already been sufficiently filtered. To this end, the first vector of $\hat{V}$ to be filtered and passed to the HEMM function is the one indicated with $s$ in `line 9` of Alg. 4. Due to space constraints, we omit the rest of the implementation details. Casting the other methods from Listing 1 into appropriate BLAS and LAPACK calls yields a complete version of ChASE.

## 5.2 Available versions of ChASE

In the software package we present two HPC versions of ChASE, which differ both in the parallelization approach and the underlying linear algebra kernels. First, we describe ChASE-MPI with a custom implementation of a distributed matrix-matrix product at its core. Then we shortly describe ChASE-Elemental, where the all linear algebra kernels are implemented in terms of the corresponding Elemental library routines.

*5.2.1 ChASE-MPI: ChASE with a custom MPI HEMM.* In ChASE the matrix operator $A$ is invoked only via matrix-matrix multiplications with (a subset of) $\hat{V}$. These HEMMs occur in the FILTER (Alg. 4, `line 9`), the Rayleigh-Ritz procedure (Alg. 2, `line 2`), and the computation of the residuals. We illustrate a version of ChASE that implements an MPI parallel HEMM. The rest of the ChASE-MPI derived class is similar to the `ChaseBLAS` implementation[5]; all non-HEMM operations on $\hat{V}$ are performed redundantly on each node using threaded BLAS. This approach is simple, flexible, and surprisingly effective. We discuss performance aspects of ChASE-MPI in Section 6.

The custom implementation of the parallel HEMM is of particular interest. Each MPI rank is assigned a block of $A$, see Fig. 3a. Such an approach requires only communication along the "rows of blocks", e.g. between MPI ranks 1 and 2, 3 and 4, etc. In general, the data distribution of $\hat{V}$ and $\hat{W}$ are different. However, the Chebyshev filter requires the alternating multiplication of $\hat{V}$ and $\hat{W}$ with

---

[5]We can interpret `ChaseBLAS` as a special case of `ChASE-MPI` for exactly one MPI rank.

Table 1. Overview of ChASE configuration parameters. Values in parentheses indicate default values for single precision.

| Category | Parameter | `ChaseConfig` member routine | Default value |
|---|---|---|---|
| General<br>see Alg. 1 | N | *constructor* | N/A |
| | nev | *constructor* | N/A |
| | nex | *constructor* | N/A |
| | approx | **void** SetApprox(**bool**); | false |
| | tol | **void** SetTol(**double**); | 1e-10 (1e-5) |
| | maxIter | **void** SetMaxIter(**size_t**); | 25 |
| Chebyshev Filter<br>see Alg. 5 and 4 | optim | **void** SetOpt(**bool**); | false |
| | degExtra | **void** SetDegExtra(**size_t**); | 2 |
| | degMax | **void** SetMaxDeg(**size_t**); | 36 (18) |
| | deg | **void** SetDeg(**size_t**); | 20 (10) |
| Spectral Estimates<br>see Alg. 6 | $k$ | **void** SetLanczosIter(**size_t**); | 25(12) |
| | $n_{\text{vec}}$ | **void** SetNumLanczos(**size_t**); | 4 |

*A*. Consequently the approach of Fig. 3a requires a re-distribution of data after each multiplication. We can avoid the re-distribution by multiplying $\hat{W}$ with the conjugate transpose of $A$, which is possible since $A$ is Hermitian, see Fig. 3b. Since all other operations are performed redundantly on each node, the full vectors have to be re-assembled via a "gather" operation. In addition to avoiding communication, the resulting matrix-matrix multiplications on each node are large and contiguous, often resulting in performance close to the theoretical peak. This data distribution easily allows the offloading of the multiplication to accelerators. ChASE currently supports a single GPU per MPI rank to accelerate the matrix-matrix multiplication. The main disadvantage of this approach is that it implements a general matrix-matrix multiply `GEMM` on each MPI rank, not an Hermitian one.

*5.2.2   ChASE-Elemental: a ChASE version with Elemental.* Elemental [35] is an excellent library for distributed-memory dense linear algebra. From an implementation point of view ChASE-Elemental is straightforward as the Elemental's kernels match ChASE's requirements very closely. Due to space constraints, we omit further details on ChASE-Elemental and point the interested reader to the source code. ChASE-Elemental has a key advantage over ChASE-MPI: it distributes $\hat{V}$ and $\hat{W}$ over the MPI processes and accordingly also parallelizes over MPI operations such as the re-orthogonalization. We discuss the performance aspect of this implementation together with ChASE-MPI in Section 6. The ChASE-Elemental version serves as an example to showcase that ChASE can easily accommodate kernels form external numerical linear algebra libraries. For example, a ChASE version based on ScaLAPACK routines would be just as straightforward.

## 5.3   ChASE Configuration Parameters

The specifics of ChASE's parameters have already been discussed in previous sections as part of the description for each algorithm. This section summarizes all of them and provides an example of how to configure and call ChASE in a code snippet.

There are three categories of configuration parameters: general parameters, parameter concerning the Chebyshev filter, and parameters for the spectral estimates. Most of the *general* parameters were introduced in Alg. 1 in Sec. 3.2. The parameters for the *Chebyshev filter* are discussed as part of Alg. 5 and Alg. 4 in Sec. 4.1, whereas the *spectral estimate* parameters concern Alg. 6 and are illustrated in Sec. 4.2. All parameters are summarized in Tab. 1. Within the source code the configuration is encapsulated by the `ChaseConfig` class. In addition to the parameter name, we specify the member function of `ChaseConfig` that sets each parameter. Further, we list the default value for each parameter; wherever applicable the values in round brackets refer to the default values for single precision data types.

Tab. 1 lists two extra parameters that we have not discussed so far: maxIter and nex. To ensure the termination of ChASE, the number of **while** loop iterations are limited to maxIter (with a default of 25) after which ChASE returns the current approximate subspace. nex specifies the size by which the search space $\hat{V}$ is incremented. In many cases increasing nex improves convergence of the SI at the cost of more expensive iteration steps. Unfortunately, the impact of nex on ChASE's performance is difficult to predict. Therefore, instead of providing a default value, we recommend an nex value between 10% and 30% of nev.

*5.3.1 Using ChASE: A code example.* Usage of ChASE amounts to constructing a valid instance of a class derived from `Chase` and then calling the `chase::Solve()` function on it. How this subclass of `ChASE` is constructed depends on the class implementation. In some cases the implementation is complicated by the data distribution; for example ChASE-MPI distributes the matrix $A$ among MPI ranks in a highly customized fashion. Usage of `ChaseBLAS`, the ChASE variant that we sketched out earlier in this section, is shown in Listing 3. We begin by setting up a `ChaseBLAS` object via the constructor defined in Listing 1. `Line 5` sets the desired residual tolerance to $10^{-10}$. Finally, we invoke the ChASE algorithm on the constructed object.

---

**Listing 3** How to call ChaseBLAS

```
1  using T = std::complex<double>;
2  void chase_driver(size_t N, size_t nev, size_t nex, T *A, T *V, T *W,
3                    double *EVs, double *resid) {
4    ChaseBlas<T> chase(N, nev, nex, A, V, W, resid);
5    single.GetConfig().SetTol(1e-10); // Set residual tolerance
6    chase::Solve(&single); // Solve for the nev smallest eigenpairs of A
7
8    cout << "Smallest Eigenvalue: " << EVs[0] << " with residual: " << resid[0] << '\n'
9  }
```

---

## 6 NUMERICAL EXPERIMENTS: SEQUENCES AND PARALLELISM

In this section we discuss performance aspects of ChASE and highlight its potential as a high performance solver. Section 6.1 deals with sequences of eigenvalue problems $\{P\}_N$. Besides using the solution $\hat{V}^{(\ell-1)}$ as the input set of vectors $\hat{V}$ when solving for the problem $P^{(\ell)}$, ChASE achieves significant performance improvements over direct solvers by varying the required accuracy for the eigenpairs residuals. Orthogonality between computed eigenvectors is inherited by the direct solver used in the Rayleigh-Ritz step (see Alg. 2, `line 3`). Strong- and weak-scaling behavior of ChASE-MPI and ChASE-Elemental are illustrated in Sec. 6.2.

For sequences of eigenvalue problems [14] a natural comparison is the Locally Optimal Block Preconditioned Conjugate Gradient (LOBPCG) method. Among the publicly available LOBPCG implementations that support dense complex-valued matrices we opt for Anasazi's [2] LOBPCG implementation. Due to reasons we discuss in Sec. 6.1 Anasazi is not well suited for runtime comparisons. Instead we compare runtimes against Elemental's state-of-the-art direct solver which uses PMRRR [33] for the tridiagonal solve. While many other distributed memory direct solvers for dense matrices exist [8, 24, 31], the performance differences between them do not significantly

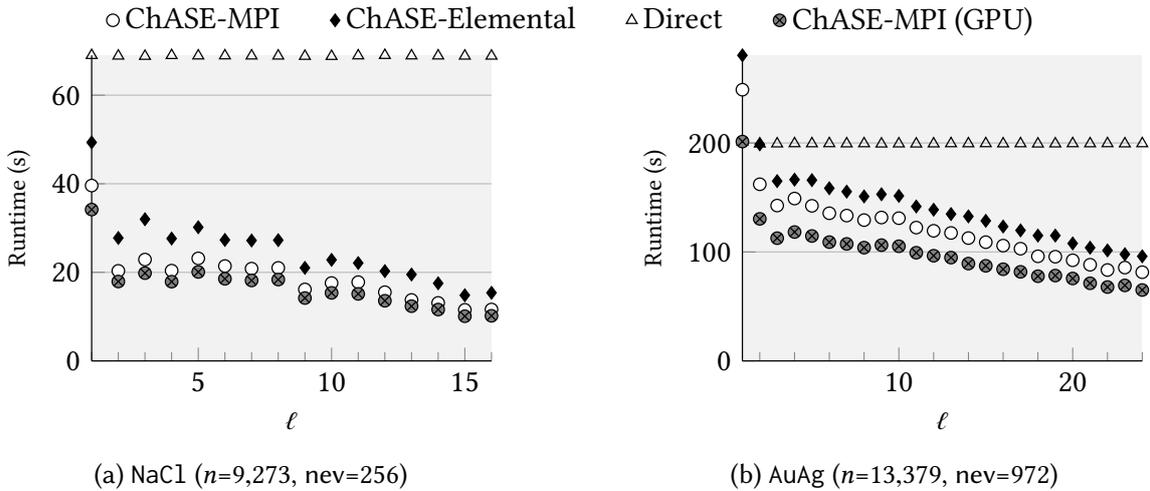

(a) NaCl ($n$=9,273, nev=256)                    (b) AuAg ($n$=13,379, nev=972)

Fig. 4. Runtimes for sequences of eigenvalue problems (single node)

contribute to our discussion. Consequently, we use Elemental's direct solver as a representative for all of them.

In our numerical tests we exclusively consider complex-valued double precision Hermitian matrices. Experiments were run on the JURECA commodity cluster located at the Jülich Supercomputing Centre. Each node of this cluster is equipped with two Xeon E5-2680 v3 Haswell CPUs over a Fat-Tree EDR Infiniband interconnect. ChASE-MPI is configured to use 48 threads per node (using HyperThreading). All software was compiled with Intel's compiler suite version 18, ParaStation MPI version 5.2.0, CUDA version 9.1. Currently, ChASE supports the use of a single GPU device per MPI rank. GPU experiments on JURECA uses half of the two available Nvidia K80's, while on BLUEWATERS ChASE makes use of the Nvidia K20 of the Cray "XK" compute nodes. Each "XK" node is equipped with an AMD 6276 Interlagos process connected via the Gemini interconnect. On BLUEWATERS ChASE is compiled with GCC v6.3.0 and linked against Cray's LibSci v16.11.1, MPICH v7.5, CUDA version 7.5. Both Elemental's direct solver and ChASE-Elemental employ the Elemental library version 0.84-p1, with 24 MPI ranks per node. Elemental distributes its matrices according to a "grid", which is essentially a two-dimensional Cartesian MPI communicator. The shape and size of the grid is critical in order to obtain good performance. As ChASE-Elemental and the Elemental direct solver usually obtain the best performance on grids with different configurations, we report the timings for the grid shape that is optimal for each solver. The reduction to tridiagonal form uses the default algorithmic block size of 64. ChASE uses the standard setting as discussed in Sec. 5.3, including a convergence criterion of tol = $10^{-10}$. The increase of the subspace (nex parameter) is chosen as 20% of nev for every problem.

## 6.1 Sequences

As we already mentioned in Sec. 2.2.2, the pivotal advantage of algorithms based on SI is the ability to use the matrix of vectors $\hat{V}$ as an approximate solution, thus effectively reducing runtime. ChASE performs particularly well because the Chebyshev accelerator reduces the number of required iterations and can be implemented in terms of HEMM calls.

Figure 4 shows runtimes for two sequences of eigenvalue problems solved on a single node using ChASE-MPI (with and without GPU), ChASE-Elemental, and Elemental's direct solver. Figure 4a shows results for the NaCl sequence of rank 9,273 and length 16, a larger version of the eigenproblem matrix seen in Section 4.1.4. Figure 4b illustrates a sequence from the new AuAg data set. As for NaCl, the AuAg sequence originates from simulations based on a DFT self-consistent field iteration based on the FLAPW method [25, 49]. The eigenproblems in AuAg have a larger matrix size $n = 13,379$, a larger sought after spectrum nev = 972, and are part of a sequence of length $N = 24$. In contrast to the NaCl data set, these matrices do not have a spectral gap, resulting in a smaller convergence ratio $|\rho_a|^{-1}$ for the eigenvalues closer to the upper end of the search space.

Table 2. Comparison of Anasazi's LOBPCG with ChASE. LOBPCG timings are not representative.

| | $\ell$ index | ChASE | | | LOBPCG | | |
|---|---|---|---|---|---|---|---|
| | | Matvecs | Iterations | Time | Matvecs | Iterations | Time |
| NaCl | 1 | 32,062 | 7 | 39.6 s | 34,153 | 76 | 232.4 s |
| | 16 | 9,636 | 2 | 11.6 s | 20,090 | 35 | 134.3 s |
| AuAg | 1 | 97,766 | 7 | 248.9 s | 99,110 | 50 | 2,493.0 s |
| | 24 | 31,052 | 2 | 81.4 s | 65,296 | 27 | 1,563.9 s |

ChASE exhibits a similar pattern for both sequences. The time to solution for both ChASE variants is much higher for the first eigenproblem of the sequence than for later ones. In the NaCl sequence, ChASE-MPI (ChASE-Elemental) requires 39 (49) seconds to converge for $\ell = 1$ and only 20 (27) seconds for $\ell = 2$, which corresponds to a speedup of 2.0 (1.8). A small part of this speedup is due to cheaper calculation of the spectral bounds (see Section 4.2), but the vast majority is caused by better starting vectors $\hat{V}$. What is more, the runtime for ChASE reduces further for increasing $\ell$ indices. On the last problem of the NaCl sequence, ChASE-MPI (ChASE-Elemental) requires only 11.6 (15.4) seconds, a total speedup over the first problem in the sequence of 3.3 (3.2). This phenomenon is due to the convergence of the underlying non-linear eigenvalue problem (see Sec. 2.2.1). The AuAg displays similar speedups. Some dips in runtime occur within a sequence for both version, e.g. at $\ell = 9$ in Figure 4a, due to natural oscillations of the angle between subspaces of adjacent DFT iterations. For all $\ell$ indices, ChASE-MPI is consistently faster than ChASE-Elemental. Profiling indicates that for this particular set of problems the matrix-matrix multiplication routine of Elemental exhibits inferior performance, probably caused by the inferior node-level parallelism of its pure MPI distribution.

The GPU version of ChASE-MPI uses the GPU for the matrix-matrix multiplication with $A$. In this manner the Chebyshev filter, the computation of the Residuals, and the Rayleigh-Ritz procedure are accelerated via the GPU. $A$ is copied to the device at the first **while** loop iteration and then reused throughout the following iterations. Only the vectors $\hat{V}$ and $\hat{W}$ are copied to and from the device. When using the GPU ChASE-MPI out-performs the non-GPU version of ChASE-MPI on all problems. Leveraging the GPU results in speedup of $\approx 1.24$ for the entire AuAg sequence. The results are slightly worse for NaCl because the smaller value for nev= 256 results in less efficient HEMMs. The GPU version achieves a speedup of respectively 1.2 and 1.1 for $\ell = 1$ and $\ell = 16$ over ChASE-MPI without GPU. The uncharacteristic speedups are due to the fact that the GK210 chip on the JURECA nodes have a peak performance ranging from 932 to 1456 GFLOPs, which is only slightly faster than the 960 GFLOPs of the CPU.

As expected, the direct solver requires the same amount of time for all eigenproblems in a sequence. However, the performance difference between ChASE and the direct solver is sequence dependent. ChASE is faster than the direct solver for all eigenproblems in the NaCl sequence. For AuAg on the other hand, ChASE is significantly slower than the direct solver for the first problem, but is faster as early as $\ell \geq 2$. This behavior is in part due to the spectral differences between the two problems (presence vs absence of a spectral gap located between nev and nev+nex for the NaCl and AuAg respectively), and in part to the different ratio $\frac{\text{nev}}{n}$ (2.8 % vs 7.3%). Of course, ChASE can only compete against a direct solver when nev—the number of eigenvalues sought—is a relatively small fraction of n [6]. A choice of nev close to n is inadvisable since the Rayleigh-Ritz subroutines contains a direct solve of rank nev.

Table 3. Matrices used in scaling experiments

| | $n$ | nev | nex | # Nodes | Jureca | | BlueWaters | |
| | | | | | # Cores | $\frac{n^2}{\text{\# Cores}}$ | # Cores | $\frac{n^2}{\text{\# Cores}}$ |
|---|---|---|---|---|---|---|---|---|
| NaCl | 3,893 | 256 | 51 | 4 | 96 | N/A | N/A | N/A |
| | 9,273 | 256 | 51 | 25 | 600 | N/A | N/A | N/A |
| AuAg | 13,379 | 972 | 194 | 25 | 600 | N/A | N/A | N/A |
| BSE | 22,360 | 100 | 20 | 9 | 216 | 2,314,674 | 64 | 7,812,025 |
| | 32,976 | 100 | 20 | 16 | 384 | 2,831,814 | 128 | 8,495,442 |
| | 47,349 | 100 | 20 | 36 | 864 | 2,594,823 | 288 | 7,784,471 |
| | 62,681 | 100 | 20 | 64 | 1,536 | 2,557,882 | 512 | 7,673,648 |
| | 76,674 | 100 | 20 | 100 | 2,400 | 2,449,542 | 800 | 7,348,628 |

As shown in [14], another iterative solver that can take advantage of an approximate solution is LOBPCG. We compare against Anasazi's LOBPCG implementation based on Trilinos version 12.12.1, with the same block size (nev + nex) and convergence tolerance of $10^{-10}$ used for ChASE. In order to match ChASE, the only non-standard setting is the usage of an absolute convergence criterion, instead of a relative one. As part of the default configuration Anasazi fully orthogonalizes the approximate eigenvectors, and locks with a tolerance of $10^{-11}$, a factor 10 lower than convergence.

Anasazi can not easily make use of all 24 cores of a Jureca node, since many of the operations on the approximate eigenvectors are bandwidth-bound. For the reasons mentioned above, a runtime comparison is not appropriate; instead we compare the number of matvecs[6]. Considering only the matvecs disregards operations on the approximate eigenvectors, such as orthogonalization, which strongly favors LOBPCG since the algorithm performs many more iterations and thus more re-orthogonalizations. In Tab. 2 we compare ChASE against LOBPCG: The table contains the number of matvecs, the number of iterations needed to achieve convergence, and the required runtime on a single Jureca node. We present data for the first and the last problem of the NaCl and the AuAg problems. The table shows that both ChASE and LOBPCG employ fewer iterations and matvecs for the last problem in the sequence that for the first one. For the first index $\ell = 1$ LOPBCG performs only slightly more matvecs that ChASE. However, as the sequence progresses, ChASE executes fewer and fewer matvecs that LOBPCG (data not shown), ending with roughly half the number of matvecs of LOBPCG for both problems.

## 6.2 Scalability

In this section we analyze ChASE's behavior both in strong- and weak-scalability regimes and compare it to Elemental's direct solver. Since the parallel efficiency of ChASE is not influenced by the use of good approximate solution[7], in this section we focus only on "single" eigenvalue problems instead of sequences. To this end, we introduce the BSE data set with considerably larger matrices that only contains sequences of length one. Note that this puts ChASE at a disadvantage over direct solvers, since there are not spectral bounds or approximate solutions available. The BSE data set contains five matrices of increasing size from $n$=22,360 to $n$=76,674, corresponding to the same physical system, but with an increasing number of representative points in momentum space. The exact sizes are available in Table 3. These matrices are the result of the discretization of the Bethe-Salpeter equation used in Optoelectronics simulations.[8] For each matrix in the set nev = 100, and we increment the subspace size by nex=20.

---

[6]Matvec is a misnomer: Both in ChASE and LOBPCG the matrix $A$ is applied to a block of vectors, resulting in a HEMM call.
[7]Experiments in Section 6.1 already uses parallelism, since they were performed on an entire node with 24 cores.

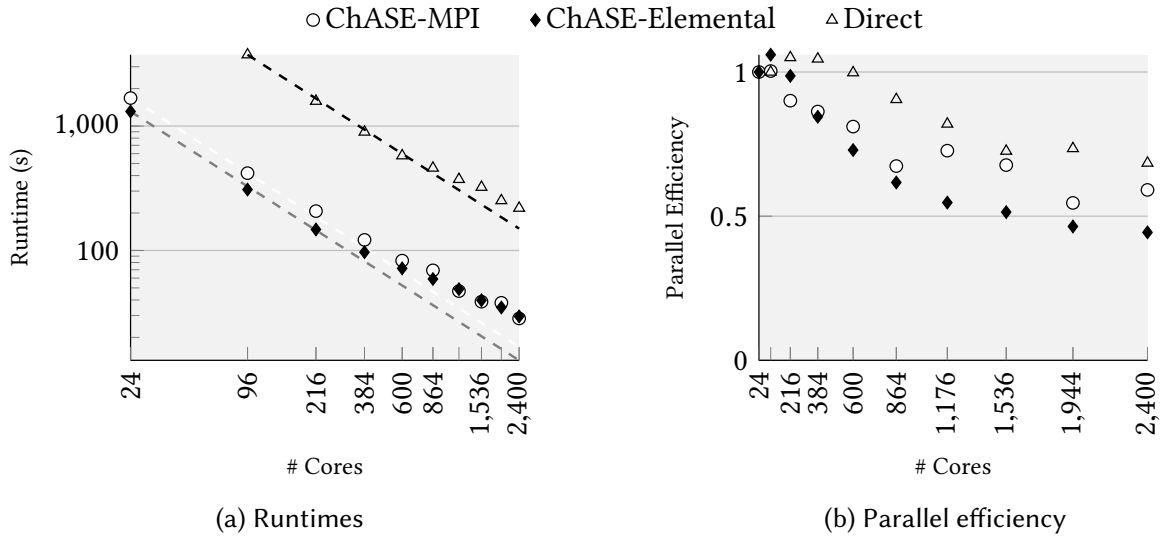

(a) Runtimes

(b) Parallel efficiency

Fig. 5. Strong scaling experiment with BSE ($n$=62,681, nev=100)

*6.2.1 Strong Scaling.* Figure 5 illustrates the results of a strong scaling experiment using the BSE matrix of size $n$=62,681. We select node counts corresponding to square numbers 9, 16, . . . , 100, as these tend to yield efficient matrix distributions, but in the plots we report the resulting number of cores (each node has 24 cores). Figure 5a reports the runtimes in a log-log plot, while parallel efficiency is shown in Figure 5b. The parallel efficiencies are calculated with respect to the timing of the corresponding solver on a single node using all available cores, as opposed to the (fastest available) sequential version.

The following remarks are in order: (1) The direct solver has a significantly longer runtime than ChASE. Both ChASE versions require roughly the same amount of time, with ChASE-Elemental being marginally faster. ChASE-MPI and ChASE-Elemental respectively solve the problem on 2400 cores in 28 and 29 seconds. Elemental's direct solver on the other hand requires 219 seconds on 2400 cores. (2) The direct solver is missing the data point for a single node, because the solver requires more than the available 128 GiB of memory. The parallel efficiency is adjusted to account for the missing data. Note that ChASE does not have this problem; it needs only $n \cdot (\text{nev} + \text{nex})$ elements of working memory. (3) All three solvers scale well, with the direct solver being the most efficient. For 2400 cores ChASE-MPI and ChASE-Elemental show a parallel efficiency of, respectively, 59% and 44%, performing only slightly worse than the direct solver with 64%. (4) Both the Elemental-based solvers (direct and ChASE-Elemental) show super-scalar speedups for smaller number of nodes, which may occur due to the large dependence of Elemental on the computing grid size: an increased number of MPI ranks may yield a more efficient grid and thus a super-scalar speedup.

*6.2.2 Weak Scaling.* Weak scaling experiments are particularly relevant to application scientists, who require the simulation of increasingly larger physical systems, which corresponds to bigger

---

Table 4. Weak scaling experiment results on Jureca

| # Cores | Iterations | | Matvecs | | Runtime | | |
| --- | --- | --- | --- | --- | --- | --- | --- |
| | ChASE-MPI | ChASE-Elemental | ChASE-MPI | ChASE-Elemental | ChASE-MPI | ChASE-Elemental | Direct |
| 216 | 11 | 11 | 19,990 | 20,192 | 25.1 s | 26.0 s | 81.5 s |
| 384 | 10 | 9 | 16,778 | 16,100 | 23.7 s | 24.0 s | 141.2 s |
| 864 | 17 | 11 | 23,424 | 27,506 | 39.8 s | 45.2 s | 211.1 s |
| 1,536 | 13 | 12 | 23,268 | 21,940 | 36.4 s | 41.4 s | 367.8 s |
| 2,400 | 10 | 13 | 22,614 | 21,720 | 38.4 s | 40.8 s | 380.1 s |

matrices. Unfortunately, in the case of an iterative eigensolver the notion of a weak scaling experiment is non-trivial. Therefore, we need to address how to set up weak scaling experiments with ChASE, before we discuss the results of the numerical tests.

Despite being generated essentially from the same physical system, ChASE performs very differently for matrices in the BSE data set, resulting in widely varying numbers of required iterations. We are also restricted to certain matrix sizes, making it difficult to have the same number of matrix elements per core. Nevertheless, in our experiment we strive for a large and roughly equal number of matrix elements per node. We collect timings for 5 different matrices from the BSE set, using between 9 and 100 nodes. The matrix sizes, number of cores, as well as the resulting number of matrix-elements per core are presented in Tab. 3. We cannot provide GPU timings for these problems on Jureca, as it only has 75 nodes equipped with GPUs. Instead we provide runtimes obtained from BlueWaters for both CPU and GPU. Tab. 4 reports the runtimes and the matvecs on Jureca, while Tab. 5 shows them for BlueWaters.

Before we discuss the runtimes of Tab. 4 in detail, note that ChASE-MPI and ChASE-Elemental yield a somewhat different number of iterations and matvecs. The difference is due to variations in the randomization of the initial subspace $\hat{V}$, and to variance in the numerical kernels, largely due to rounding errors in floating-point arithmetic. Tab. 4 shows that the direct solver needs much larger runtimes than ChASE. Moreover, the direct solver execution increases drastically for larger ranks, as the reduction to tridiagonal form is of cubic complexity. The runtimes for the two ChASE versions are roughly similar, and surprisingly do not strictly increase with the problem size. We observe that the number of required iterations changes just as much from one ChASE implementation as is does for the separate BSE problems. In addition, the number of matvecs and iterations does not necessarily increase with problem size. We conclude that for a realistic scaling scenario the spectral differences between the increasingly larger problems have a bigger influence on performance than any adverse weak scaling behavior of ChASE. Finally for ChASE-MPI the number of matvecs is roughly similar and so are the runtimes, implying that ChASE-MPI does scales well in a weak scaling regime.

Tab. 5 contains runtimes, matvecs, and iterations for ChASE-MPI with and without GPU. In addition to the total runtime and speedup achieved via the GPU, we also report these metrics for the Chebyshev filter. The BlueWaters "XK" node has a peak CPU performance of 156 GFLOPs on the CPU, 1310 GFLOPs on the available K20 GPU. Aside from the performance difference, the CPU runtimes behavior looks similar to the one observed on Jureca: the number of iterations changes from problem to problem, and the runtimes fluctuate accordingly. The GPU timings, however, provide new insight. On the Chebyshev filter ChASE leverages the GPU for a 6.4 – 7.7 speedup over the CPU. The rather small nev = 100 value influences the latency of the transfer from GPU to RAM—required for the parallel matrix-matrix-multiply— and is a significant source of overhead. For larger values of nev the incurred latency amortizes over more computation, resulting in better speedups.

Table 5. Weak scaling experiment results for ChASE-MPI on BlueWaters

| # Cores | Iterations | Matvecs | Filter Runtime | | | Total Runtime | | |
|---:|---:|---:|---:|---:|---:|---:|---:|---:|
| | | | CPU | GPU | Speedup | CPU | GPU | Speedup |
| 64 | 11 | 20,106 | 176.4 s | 22.8 s | 7.7 | 228.0 s | 43.5 s | 5.2 |
| 128 | 9 | 16,856 | 175.9 s | 27.5 s | 6.4 | 236.1 s | 52.6 s | 4.5 |
| 288 | 12 | 23,610 | 231.5 s | 30.2 s | 7.7 | 306.8 s | 70.3 s | 4.4 |
| 512 | 14 | 23,080 | 225.5 s | 30.1 s | 7.5 | 316.5 s | 87.3 s | 3.6 |
| 800 | 12 | 22,868 | 209.1 s | 30.8 s | 6.8 | 299.2 s | 89.9 s | 3.3 |

The speedup from the GPU over the total runtime decreases for the experiments involving larger matrix sizes. While the speedup for the smallest problem is 5.2, it decreases to 3.3 for the largest problem. Because ChASE-MPI parallelizes only the matrix-matrix multiplication of $\hat{V}$ with $A$ over MPI ranks, operations such as the QR are not parallelized but executed redundantly on the node. Thus, when the matrix size becomes bigger, the costs for the orthogonalization increase accordingly. For the largest matrix, the Chebyshev filter accounts for only a third of the runtime (30.8 s vs. 89.9 s).

## 7 CONCLUSIONS AND OUTLOOK

We presented ChASE, a modern C++ library based on subspace iteration complemented with a Chebyshev polynomial filter. ChASE targets extremal dense eigenproblems, when a relatively small fraction ($\leq 15\%$) of the eigenpairs is sought after. There are several aspects that distinguish ChASE from software projects based on similar algorithms:
— The C++ implementation features a modern and flexible interface that separates the algorithm from the data distribution and kernel execution. As such, the ChASE library is relatively easy to port to specialized computing architectures and simple to integrate in existing application codes.
— The library is particularly effective in exploiting the correlation among the solutions of eigenproblems which are part of a sequence: when inputted in ChASE, eigenvectors of $P^{(\ell-1)}$ substantially speedup the solution of the next problem $P^{(\ell)}$.
— In the absence of approximate solutions, spectral estimates are computed with an enhanced algorithm based on the computation of a spectral density using an repeated Lanczos processes.
— The polynomial degree of the Chebyshev filter is optimized for each single vector so as to minimize the number of FLOPs needed to achieve convergence. The optimization saves, on average, about 20% of the required FLOPs.
— The library comes equipped with two distinct parallelized versions, one of which is GPU-compatible.

In the case of sequences of eigenproblems, ChASE outperforms state-of-the-art dense direct eigensolvers for values of the sought after exterior spectrum equal or less than 10%. This result is valid even in the worst-case scenarios, when the absence of a spectral gap at the end of the search space may hamper the ratio of convergence of the filtered vectors. The exploitation of approximate solutions contributes in great part to the success of this result. ChASE scales well up to thousands of cores, it does so with a good parallel efficiency for both strong- and weak-scaling regimes, and is in line with the state-of-the-art scalability of a direct solver. This is due to the fact that ChASE concentrates most of its work (about 80+%) in matrix-matrix multiplications, for which it uses BLAS level 3 kernels. ChASE capitalizes on the extreme effectiveness of these kernels to achieve a high efficiency both on the node and in a parallel setting. For the same reason ChASE performs particularly well in weak scaling regimes, although the convergence behavior of each problem must be taken into account by considering the number of required matvecs. ChASE can make use of a GPU to accelerate the matrix-matrix multiplication even for values of nev as small as 100, resulting in $\geq 6.4$ speedup on a Cray "XK" node.

In the mid-term we plan to expand support for sparse matrices through SpGEMM kernels. In the near term, a version of ChASE that supports distribution of the workload on multiple GPUs per compute node will be added to the available implementation.

### Software

ChASE is open source (BSD license 3.0) and available on GitHub
https://github.com/SimLabQuantumMaterials/ChASE

### ACKNOWLEDGMENTS

The ChASE library is the final chapter in an effort that saw the involvement of many people since its initial inception. We would like to thank Y. Zhou for having initially suggested the algorithm, M. Berljafa for the first implementation of the library in C++ and its successive parallelization based on data distribution and kernels of the Elemental library, J. R. Suckert and J. Zubriniĉ for adaptation to GPUs, and J. Poulson for its support with the Elemental library usage. The numerical tests would have not been possible without the kind support of D. Wortmann in providing the physical systems from which sequences of problems were generated, and A. Schleife in supplying the matrices used in the weak scaling tests.

### REFERENCES

[1] Edward Anderson, Zhaojun Bai, Christian Bischof, L. Susan Blackford, James Demmel, Jack Dongarra, Jeremy Du Croz, Anne Greenbaum, Sven Hammarling, Alan McKenney, et al. 1999. *LAPACK Users' Guide*. SIAM, Philadelphia, PA. https://doi.org/10.1137/1.9780898719604

[2] Chris G. Baker, Ulrich L. Hetmaniuk, Richard B. Lehoucq, and Heidi K. Thornquist. 2009. Anasazi Software for the Numerical Solution of Large-scale Eigenvalue Problems. *ACM Trans. Math. Software* 36, 3, Article 13 (July 2009), 23 pages. https://doi.org/10.1145/1527286.1527287

[3] Satish Balay, Shrirang Abhyankar, Mark F. Adams, Jed Brown, Peter Brune, Kris Buschelman, Lisandro Dalcin, Victor Eijkhout, William D. Gropp, Dinesh Kaushik, Matthew G. Knepley, Dave A. May, Lois Curfman McInnes, Karl Rupp, Barry F. Smith, Stefano Zampini, Hong Zhang, and Hong Zhang. 2018. PETSc Web page. Retrieved Mar 3, 2018 from http://www.mcs.anl.gov/petsc

[4] Amartya S. Banerjee, Lin Lin, Wei Hu, Chao Yang, and John E. Pask. 2016. Chebyshev polynomial filtered subspace iteration in the discontinuous Galerkin method for large-scale electronic structure calculations. *The Journal of Chemical Physics* 145, 15 (Oct. 2016), 154101. https://doi.org/10.1063/1.4964861

[5] Friedrich L. Bauer. 1957. Das Verfahren der Treppeniteration und verwandte Verfahren zur Lösung algebraischer Eigenwertprobleme. *Zeitschrift für Angewandte Mathematik und Physik ZAMP* 8, 3 (May 1957), 214–235. https://doi.org/10.1007/BF01600502

[6] Mario Berljafa, Daniel Wortmann, and Edoardo Di Napoli. 2015. An optimized and scalable eigensolver for sequences of eigenvalue problems. *Concurrency and Computation: Practice and Experience* 27 (Sept. 2015), 905–922. https://doi.org/10.1002/cpe.3394

[7] Paolo Bientinesi, Inderjit S. Dhillon, and Robert A. van de Geijn. 2005. A Parallel Eigensolver for Dense Symmetric Matrices Based on Multiple Relatively Robust Representations. *J. Scientific Computing* 27, 1 (Jan. 2005), 43–66. https://doi.org/10.1137/030601107

[8] L. S. Blackford, J. Choi, A. Cleary, E. D'Azevedo, J. Demmel, I. Dhillon, J. Dongarra, S. Hammarling, G. Henry, A. Petitet, K. Stanley, D. Walker, and R. C. Whaley. 1997. *ScaLAPACK Users' Guide*. SIAM, Philadelphia, PA. https://doi.org/10.1137/1.9780898719642

[9] Stefan Blügel, Gustav Bihlmayer, and Daniel Wortmann. 2017. The FLEUR code. http://www.flapw.de. Accessed: 2018-01-22.

[10] Maurice Clint and A. Jenning. 1970. The evaluation of eigenvalues and eigenvectors of real symmetric matrices by simultaneous iteration. *Comput. J.* 13, 1 (1970), 76–80. https://doi.org/10.1093/comjnl/13.1.76

[11] Jan J. M. Cuppen. 1980. A divide and conquer method for the symmetric tridiagonal eigenproblem. *Numer. Math.* 36, 2 (March 1980), 177–195. https://doi.org/10.1007/BF01396757

[12] Inderjit S. Dhillon. 1997. *A New $O(n^2)$ Algorithm for the Symmetric Tridiagonal Eigenvalue/Eigenvector Problem.* Ph.D. Dissertation. EECS Department, University of California, Berkeley. http://www2.eecs.berkeley.edu/Pubs/TechRpts/1997/5888.html

[13] Edoardo Di Napoli. 2018. Chebyshev acceleration of subspace iteration revisited: convergence and optimal polynomial degree. (2018). In preparation.

[14] Edoardo Di Napoli and Mario Berljafa. 2013. Block iterative eigensolvers for sequences of correlated eigenvalue problems. *Computer Physics Communications* 184, 11 (Nov. 2013), 2478–2488. https://doi.org/10.1016/j.cpc.2013.06.017

[15] Edoardo Di Napoli, Stefan Blügel, and Paolo Bientinesi. 2012. Correlations in sequences of generalized eigenproblems arising in Density Functional Theory. *Computer Physics Communications* 183, 8 (Aug. 2012), 1674–1682. https://doi.org/10.1016/j.cpc.2012.03.006

[16] Jack Dongarra, Jeremy Du Croz, Sven Hammarling, and Iain Duff. 1990. A set of level 3 basic linear algebra subprograms. *ACM Trans. Math. Software* 16, 1 (March 1990), 1–17. https://doi.org/10.1145/77626.79170

[17] Jack Dongarra, Victor Eijkhout, and Ajay Kalhan. 1995. *Reverse Communication Interface for Linear Algebra Templates for Iterative Methods.* Technical Report. University of Tennessee, Knoxville, TN, USA.

[18] Gene H. Golub and Hongyuan Zha. 1995. The canonical correlations of matrix pairs and their numerical computation. In *Linear algebra for signal processing.* Springer, New York, 27–49. https://doi.org/10.1007/978-1-4612-4228-4_3

[19] Ming Gu and Stanley C. Eisenstat. 1995. A Divide-and-Conquer Algorithm for the Symmetrical Tridiagonal Eigenproblem. *SIAM J. Matrix Anal. Appl.* 16, 1 (Jan. 1995), 172–191. https://doi.org/10.1137/S0895479892241287

[20] Vicente Hernandez, Jose E. Roman, and Vicente Vidal. 2005. SLEPc: A Scalable and Flexible Toolkit for the Solution of Eigenvalue Problems. *ACM Trans. Math. Software* 31, 3 (Sept. 2005), 351–362. https://doi.org/10.1145/1089014.1089019

[21] Michael A. Heroux, Roscoe A. Bartlett, Vicki E. Howle, Robert J. Hoekstra, Jonathan J. Hu, Tamara G. Kolda, Richard B. Lehoucq, Kevin R. Long, Roger P. Pawlowski, Eric T. Phipps, Andrew G. Salinger, Heidi K. Thornquist, Ray S. Tuminaro, James M. Willenbring, Alan Williams, and Kendall S. Stanley. 2005. An overview of the Trilinos project. *ACM Trans. Math. Software* 31, 3 (Sept. 2005), 397–423. https://doi.org/10.1145/1089014.1089021

[22] Pierre Hohenberg and Walter Kohn. 1964. Inhomogeneous Electron Gas. *Physical Review* 136 (Nov 1964), B864–B871. Issue 3B. https://doi.org/10.1103/PhysRev.136.B864

[23] Harold Hotelling. 1936. Relations Between Two Sets of Variates. *Biometrika* 28, 3/4 (Dec. 1936), 321–377. https://doi.org/10.1007/978-1-4612-4380-9_14

[24] Toshiyuki Imamura, Susumu Yamada, and Masahiko Machida. 2011. Development of a high performance eigensolver on the petascale next generation supercomputer system. *Progress in Nuclear Science and Technology* 2 (2011), 643–650. https://doi.org/10.15669/pnst.2.643

[25] H. J. F. Jansen and A. J. Freeman. 1984. Total-Energy Full-Potential Linearized Augmented-Plane-Wave Method for Bulk Solids - Electronic and Structural-Properties of Tungsten. *Physical Review B* 30, 2 (July 1984), 561–569. https://doi.org/10.1103/PhysRevB.30.561

[26] Walter Kohn. 1999. Nobel Lecture: Electronic structure of matter-wave functions and density functionals. *Reviews of Modern Physics* 71, 5 (Oct. 1999), 1253–1266. https://doi.org/10.1103/RevModPhys.71.1253

[27] R. Lehoucq, D. Sorensen, and C. Yang. 1998. *ARPACK Users' Guide.* Vol. 6. SIAM, Philadelphia, PA. https://doi.org/10.1137/1.9780898719628

[28] Antoine Levitt and Marc Torrent. 2015. Parallel eigensolvers in plane-wave Density Functional Theory. *Computer Physics Communications* 187 (Feb. 2015), 98–105. https://doi.org/10.1016/j.cpc.2014.10.015

[29] Lin Lin, Yousef Saad, and Chao Yang. 2016. Approximating spectral densities of large matrices. *SIAM Rev.* 58, 1 (2016), 34–65. https://doi.org/10.1137/130934283

[30] S. H. Lui. 2000. Domain decomposition methods for eigenvalue problems. *J. Comput. Appl. Math.* 117, 1 (May 2000), 17–34. https://doi.org/10.1016/S0377-0427(99)00326-X

[31] Andreas Marek, Volker Blum, Rainer Johanni, Ville Havu, Bruno Lang, Thomas Auckenthaler, Alexander Heinecke, Hans-Joachim Bungartz, and Hermann Lederer. 2014. The ELPA library: scalable parallel eigenvalue solutions for electronic structure theory and computational science. *Journal of Physics: Condensed Matter* 26, 21 (2014), 213201. https://doi.org/10.1088/0953-8984/26/21/213201

[32] Beresford N. Parlett. 1998. *The Symmetric Eigenvalue Problem.* SIAM, Philadelphia, PA. https://doi.org/10.1137/1.9781611971163

[33] Matthias Petschow, Elmar Peise, and Paolo Bientinesi. 2013. High-Performance Solvers for Dense Hermitian Eigenproblems. *SIAM Journal on Scientific Computing* 35, 1 (2013), C1–C22. https://doi.org/10.1137/110848803

[34] Eric Polizzi. 2009. Density-matrix-based Algorithm for Solving Eigenvalue Problems. *Physical Review B* 79 (Mar 2009), 115112. https://doi.org/10.1103/PhysRevB.79.115112

[35] Jack Poulson, Bryan Marker, Robert A. van de Geijn, Jeff R. Hammond, and Nichols A. Romero. 2013. Elemental: A New Framework for Distributed Memory Dense Matrix Computations. *ACM Trans. Math. Software* 39, 2, Article 13 (Feb. 2013), 24 pages. https://doi.org/10.1145/2427023.2427030

[36] E. H. Rogers. 1964. A Minimax Theory for Overdamped Systems. *Archive for Rational Mechanics and Analysis* 16, 2 (1964), 89–96. https://doi.org/10.1007/bf00281333

[37] Axel Ruhe. 1973. Algorithms for the Nonlinear Eigenvalue Problem. *SIAM J. Numer. Anal.* 10, 4 (Sept. 1973), 674–689. https://doi.org/10.1137/0710059

[38] Heinz Rutishauser. 1969. Computational aspects of FL Bauer's simultaneous iteration method. *Numer. Math.* 13, 1 (1969), 4–13. https://doi.org/10.1007/BF02165269

[39] Heinz Rutishauser. 1970. Simultaneous iteration method for symmetric matrices. *Numer. Math.* 16, 3 (1970), 205–223. https://doi.org/10.1007/bf02219773

[40] Yousef Saad. 2011. *Numerical methods for large eigenvalue problems.* SIAM, Philadelphia, PA. https://doi.org/10.1137/1.9781611970739

[41] Pablo Salas, Luc Giraud, Yousef Saad, and Stéphane Moreau. 2015. Spectral recycling strategies for the solution of nonlinear eigenproblems in thermoacoustics. *Numerical Linear Algebra with Applications* 22, 6 (2015), 1039–1058. https://doi.org/10.1002/nla.1995

[42] Andreas Stathopoulos. 2006. *Locking issues for finding a large number of eigenvectors of Hermitian matrices.* Technical Report. Department of Computer Science, College of William and Mary, Williamsburg,. Virginia 23187-8795. Tech Report: WM-CS-2005-09.

[43] Andreas Stathopoulos and James R. McCombs. 2010. PRIMME: preconditioned iterative multimethod eigensolver—methods and software description. *ACM Trans. Math. Software* 37, 2 (2010), 21. https://doi.org/10.1145/1731022.1731031

[44] G. W. Stewart. 1969. Accelerating the orthogonal iteration for the eigenvectors of a Hermitian matrix. *Numer. Math.* 13, 4 (Aug. 1969), 362–376. https://doi.org/10.1007/BF02165413

[45] G. W. Stewart. 1976. Simultaneous iteration for computing invariant subspaces of non-Hermitian matrices. *Numer. Math.* 25, 1 (1976), 123–136. https://doi.org/10.1007/BF01462265

[46] G. W. Stewart. 2001. *Matrix algorithms. Vol. II.* SIAM, Philadelphia, PA. https://doi.org/10.1137/1.9780898718058

[47] Christof Vömel, Ratko G. Veprek, Urban Weber, and Peter Arbenz. 2012. *Iterative solution of generalized eigenvalue problems from optoelectronics with trilinos.* Technical Report. Computer Science Department, ETH Zürich. 15 pages. https://doi.org/10.3929/ethz-a-006821768

[48] Heinrich Voss and Bodo Werner. 1982. A minimax principle for nonlinear eigenvalue problems with applications to nonoverdamped systems. *Mathematical Methods in the Applied Sciences* 4 (1982), 415–424. https://doi.org/10.1002/mma.1670040126

[49] Erich Wimmer, Henry Krakauer, Michael Weinert, and A. J. Freeman. 1981. Full-potential self-consistent linearized-augmented-plane-wave method for calculating the electronic-structure of molecules and surfaces: O2 Molecule. *Physical Review B* 24, 2 (July 1981), 864–875. https://doi.org/10.1103/PhysRevB.24.864

[50] T. Zhang, Gene H. Golub, and Kincho H. Law. 1999. Subspace iterative methods for eigenvalue problems. *Linear Algebra Appl.* 294, 1 (1999), 239 – 258. https://doi.org/10.1016/S0024-3795(99)00074-9

[51] Yunkai Zhou, James R. Chelikowsky, and Yousef Saad. 2014. Chebyshev-filtered subspace iteration method free of sparse diagonalization for solving the Kohn–Sham equation. *J. Comput. Phys.* 274 (Oct. 2014), 770–782.

[52] Yunkai Zhou and Ren-Cang Li. 2011. Bounding the spectrum of large Hermitian matrices. *Linear Algebra Appl.* 435, 3 (Aug. 2011), 480–493.

[53] Yunkai Zhou, Yousef Saad, Murilo L. Tiago, and James R. Chelikowsky. 2006. Self-consistent-field calculations using Chebyshev-filtered subspace iteration. *J. Comput. Phys.* 219, 1 (Nov. 2006), 172–184. https://doi.org/10.1016/j.jcp.2006.03.017